\documentclass{emulateapj}

\newcommand{\sings}{SINGS}
\newcommand{\spitzer}{{\em Spitzer}}
\newcommand{\galex}{GALEX}
\newcommand{\things}{THINGS}
\newcommand{\bima}{BIMA SONG}
\newcommand{\hi}{\mbox{\rm H\,{\sc i}}}
\newcommand{\hii}{\mbox{\rm H\,{\sc ii}}}
\newcommand{\htwo}{\mbox{\rm H$_2$}}

\newcommand{\halpha}{\mbox{\rm H$\alpha$}}
\newcommand{\microm}{\micron}

\newcommand{\xcounits}{\mbox{cm$^{-2}$ (K km s$^{-1}$)$^{-1}$}}

\shorttitle{The Star Formation Law in Nearby Galaxies on Sub-Kpc Scales}
\shortauthors{F. Bigiel et al.}

\begin{document}
\slugcomment{Accepted for Publication in The Astronomical Journal}
\title{The Star Formation Law in Nearby Galaxies on Sub-Kpc Scales}
\author{F.~Bigiel\altaffilmark{1}, A.~Leroy\altaffilmark{1},
  F.~Walter\altaffilmark{1}, E.~Brinks\altaffilmark{2},
  W.~J.~G.~de~Blok\altaffilmark{3}, B.~Madore\altaffilmark{4},
  M.~D.~Thornley\altaffilmark{5}}

\altaffiltext{1}{Max-Planck-Institut f{\"u}r Astronomie, K{\"o}nigstuhl 17,
  69117, Heidelberg, Germany; bigiel@mpia.de}
\altaffiltext{2}{Centre for Astrophysics Research, University of
  Hertfordshire, Hatfield AL10 9AB, U.K.}
\altaffiltext{3}{Department of Astronomy, University of Cape Town, Private Bag
  X3, Rondebosch 7701, South Africa}
\altaffiltext{4}{Observatories of the Carnegie Institution of Washington,
  Pasadena, CA 91101, USA}
\altaffiltext{5}{Department of Physics and Astronomy, Bucknell
  University, Lewisburg, PA 17837, USA}

\begin{abstract}

  We present a comprehensive analysis of the relationship between star
  formation rate surface density, $\Sigma_{\rm SFR}$, and gas surface density, $\Sigma_{\rm gas}$,
  at sub--kpc resolution in a sample of 18 nearby
  galaxies. We use high resolution \hi\ data from \things,
  CO data from HERACLES and BIMA SONG,
  24\,\microm\ data from the \spitzer\ Space Telescope, and UV data from
  \galex. We target 7 spiral galaxies and 11
  late-type/dwarf galaxies and investigate how the star formation law
  differs between the H$_2$-dominated centers of spiral galaxies,
  their \hi-dominated outskirts and the \hi-rich late-type/dwarf
  galaxies. We find that a Schmidt-type power law with index $N=1.0 \pm
  0.2$ relates $\Sigma_{\rm SFR}$ and $\Sigma_{\rm H2}$ across our
  sample of spiral galaxies, i.e., that H$_2$ forms stars at a
  constant efficiency in spirals. The average molecular gas depletion time
  is $\sim2 \cdot 10^9$ years. The range of $\Sigma_{\rm H2}$ over
  which we measure this relation is $\sim3-50$~M$_{\odot}$~pc$^{-2}$, significantly
  lower than in starburst environments. We find the same results when
  performing a pixel-by-pixel analysis, averaging in radial bins, or when
  varying the star formation tracer used. We interpret
  the linear relation and constant depletion time as evidence that
  stars are forming in GMCs with approximately uniform properties and that
  $\Sigma_{\rm H2}$ may be more a measure
  of the filling fraction of giant molecular clouds than changing
  conditions in the molecular gas. The relationship
  between total gas surface density ($\Sigma_{\rm gas}$) and $\Sigma_{\rm SFR}$
  varies dramatically among and within spiral galaxies. Most galaxies show little or no
  correlation between $\Sigma_{\rm HI}$ and $\Sigma_{\rm SFR}$. As a
  result, the star formation efficiency (SFE), $\Sigma_{\rm
    SFR}/\Sigma_{\rm gas}$, varies strongly across our sample and within
  individual galaxies. We show
  that this variation is systematic and consistent with the SFE being set
  by local environmental factors: in spirals the SFE is a clear
  function of radius, while the dwarf galaxies in our sample display
  SFEs similar to those found in the outer optical disks of the spirals.
  We attribute the similarity to common
  environments (low-density, low-metallicity, \hi-dominated) and
  argue that shear (which is typically absent in dwarfs) cannot drive the SFE.
  In addition to a molecular Schmidt law, the other general feature of
  our sample is a sharp saturation of \hi\ surface densities at
  $\Sigma_{\rm HI} \approx 9$~M$_{\odot}$~pc$^{-2}$ in both the
  spiral and dwarf galaxies. In the case of the spirals, we observe gas in
  excess of this limit to be molecular.

\end{abstract}


\section{Introduction}
\label{intro}

A robust, quantitative measurement of the relationship between star
formation rate (SFR) and gas density (SF law) is of major astrophysical importance in
the context of galaxy evolution: it describes how efficiently galaxies turn their
gas into stars and constrains theoretical models of
star formation; in addition, it serves as essential input to simulations and models
of galaxy evolution
\citep[e.g.,][]{springel03,boissier99,tan99,krumholz05,matteucci06}.
Direct observations of this relationship at sub-kpc scales are
still very rare because measuring the distributions of star formation,
atomic gas, and molecular gas at high resolution and sensitivity is
challenging and time-consuming.

However, the past few years have seen an explosion
in multiwavelength data for nearby galaxies. From the `GALEX Nearby
Galaxy Survey' \citep{gildepaz07} and the `{\em Spitzer} Infrared Nearby
Galaxies Survey' \citep[SINGS,][]{KENNICUTT03} the distribution of star
formation is now known in a large suite of local galaxies. From the
`BIMA Survey of Nearby Galaxies' \citep[\bima,][]{helfer03} and recent
observations with the IRAM 30m telescope \citep[HERACLES,][]{leroy08-2} the CO
distributions of many nearby galaxies are also known. With `The HI
Nearby Galaxy Survey' \citep[THINGS,][]{WALTER08}, \hi\ maps that match or exceed the angular resolution and sensitivity of the ultraviolet, infrared, and CO data are now available for 34 nearby
galaxies. In this paper, we combine this suite of multiwavelength
data to measure the surface densities of \hi, H$_2$, and SFR
($\Sigma_{\rm HI}$, $\Sigma_{\rm H2}$, and $\Sigma_{\rm SFR}$) across
the entire optical disks of 18 nearby galaxies at 750~pc spatial
resolution. Using these measurements, we examine the relationships among these three
quantities across our sample.

Following the pioneering work of \citet{SCHMIDT59}, it is common to
relate gas density to the SFR density using a power law.
He suggested the form $\rho_{\rm
  SFR}\,\sim\,\left(\rho_{\rm gas}\right)^{n}$, where $\rho_{\rm
  SFR}$ and $\rho_{\rm gas}$ denote the volume densities of
the SFR and the gas.  Studying the distribution of \hi\ and stars
perpendicular to the Galactic plane, he derived a power-law index of
$n\approx2$. \citet{sanduleak69} and \citet{hartwick71} carried out the first
measurements of the Schmidt law in other galaxies. They used
bright stars in the Small Magellanic Cloud and \hii\ regions in M31, respectively,
to trace star formation and focused their analyses on \textit{surface}
densities, $\Sigma_{\rm HI}$ and $\Sigma_{\rm SFR}$ , which are
directly observable (i.e., $\Sigma_{\rm SFR}\,\sim\,\left(\Sigma_{\rm gas}\right)^{N}$). They found $N_{\rm SMC} = 1.84 \pm 0.14$ and $N_{\rm M31} = 3.50 \pm 0.12$. Throughout this paper, we use the term Schmidt law or star formation law (SF law) to refer to a power law formalism relating star formation surface density to atomic, molecular, or total gas surface density. Note that for a constant scale height, the exponents $N$ and $n$ are identical, i.e., it does not matter whether one considers volume or surface densities.

\citet{madore74} compared stars and \hii-regions to $\Sigma_{\rm HI}$ and
derived a higher power law index for the outer part than for the inner part of M33. \citet{newton80} repeated this analysis with higher-resolution \hi\ data and confirmed this behavior. \citet{tosa75} compared \hii-regions to $\Sigma_{\rm HI}$ in M31 and the LMC and find $N\approx2$ for both galaxies. \citet{hamajima75} performed the same analysis in 7 nearby galaxies. They derived power law indices $N=1.5-2.9$ and found radial variations in $N$ for 2 of their sample galaxies (M31 and M101).

\citet[][hereafter K98]{KENNICUTT89,KENNICUTT98} studied
the globally averaged relationship between SFR and gas in a sample of 61 nearby normal
spiral and 36 infrared-selected starburst galaxies. K98 showed that a
Schmidt law relates the disk-averaged total gas surface density,
$\Sigma_{\rm gas} =\Sigma_{\rm HI} + \Sigma_{\rm H2}$, to the
disk-averaged star formation surface density, $\Sigma_{\rm SFR}$, over
many orders of magnitude. His subsample of normal spiral galaxies
yields a power-law index $N=2.47 \pm 0.39$; his composite sample, including
starburst galaxies, yields $N=1.40 \pm 0.15$. Similar studies of the
disk-averaged Schmidt Law used a range of SFR tracers --
such as \halpha, UV, radio continuum, and FIR emission -- and found 
$N=0.9-1.7$ \citep[e.g.,][]{buat89,buat92,deharveng94}.

Other authors studied a local Schmidt law using radial profiles (i.e., comparing
azimuthally averaged values) of $\Sigma_{\rm SFR}$ and $\Sigma_{\rm gas}$.
\citet{wong02} found $N = 1.2$ -- $2.1$ for 6 molecule-rich
spiral galaxies. \citet{boissier03} derived $N\,\approx 2$ for 16
galaxies and \citet{misiriotis06} obtained $N = 2.18 \pm 0.20$ for the
Milky Way. \citet{heyer04} found $N = 3.3 \pm 0.1$ for M33. For M51,
\citet{schuster07} found $N = 1.4 \pm 0.6$ and \citet{KENNICUTT07}, using
520\,pc apertures centered on \halpha\ and 24\,\microm\ emission peaks, found $N = 1.56 \pm 0.04$.

The large range of power-law indices in the literature,
$N \approx 1$--$3$, suggests that
either different SF laws exist in different galaxies or that $N$ is very
sensitive to systematic differences in methodology (e.g., the choice
of SFR tracers, the spatial resolution of the data, etc.). There are
physically motivated, theoretical reasons to expect $N$ in the range $0.75$ -- $2$, and the
precise value of $N$ may vary with the regime one considers.
\citet{krumholz05} argue that a giant molecular cloud (GMC) will
convert its gas into stars over a free-fall time, $t_{\rm ff} \propto
\rho_{\rm gas}^{-0.5}$.  If this is the case, then from the
observation that the surface density of GMCs is constant
\citep[e.g.,][]{solomon87,blitz07}, implying $\rho_{\rm gas} \propto M_{\rm
GMC}^{-0.5}$, one expects $N \approx 0.75$ for the
case where we measure the Schmidt law for individual GMCs. If we
instead compare uniform populations of GMCs, where the different
molecular gas surface densities reflect only different {\em numbers}
of clouds (not different physical properties), then one expects $N \approx 1$ for the
molecular gas exponent (assuming the population-averaged timescale over which a
GMC converts its gas into stars is constant). If gravitational
instability in the neutral (\hi+H$_2$) ISM is the key process in
star formation, the SFR may instead depend on the free-fall time of
the {\em total} gas; in this case, and if the gas scale height is constant,
$t_{\rm ff} \propto \Sigma_{\rm gas}^{-0.5}$ and one would expect $N
\approx 1.5$ \citep[e.g.,][]{madore77}. Finally, if one postulates that star formation is a
fundamentally collisional process, because, e.g., it may depend on the
formation of H$_2$ by collisions between hydrogen atoms and dust
grains or the collision of small clouds to form larger clouds, one
would expect $\rho_{\rm SFR} \propto \rho_{\rm gas}^2$, which will
lead to $N \approx 2$ (again for a system with constant gas scale height).

Another open question is whether the Schmidt law is fundamentally a
molecular phenomenon or if a single, universal power law relates total
gas and star formation. Because all stars are believed to form in
molecular clouds, it would seem natural that H$_2$ and SFR are more
immediately related than \hi\ or total gas and SFR. Therefore it is
somewhat surprising that observations remain contradictory on this
point. One of the most surprising findings by K98 was that
$\Sigma_{\rm SFR}$ correlated better with $\Sigma_{\rm HI}$ than with
$\Sigma_{\rm H2}$ in normal disks; the strongest correlation, between
$\Sigma_{\rm SFR}$ and $\Sigma_{\rm gas}$, was only marginally
stronger than the $\Sigma_{\rm SFR}$ -- $\Sigma_{\rm HI}$ correlation.
\citet{wong02} found a much stronger relationship between $\Sigma_{\rm
  SFR}$ and $\Sigma_{\rm H2}$ than between $\Sigma_{\rm SFR}$ and
$\Sigma_{\rm HI}$, even finding an anti-correlation between
$\Sigma_{\rm SFR}$ and $\Sigma_{\rm HI}$ at high SFRs, but they
focused on molecule-rich spirals and therefore did not include large
amounts of the \hi-dominated ISM. Even in H$_{2}$-rich spirals, the
question is still open. Recently, \citet{schuster07} and
\citet{crosthwaite07} studied the molecular gas-rich spirals NGC~5194 and
NGC~6946 and found that the total gas correlates
better with the SFR than H$_2$ alone.

This paper attempts to answer the following questions: how do the SFR, H$_2$, and
\hi\ surface densities relate to one another in nearby galaxies on a pixel-by-pixel basis?
Which of these relationships are common across our sample, and which vary with environment? To
address these questions, we measure $\Sigma_{\rm HI}$, $\Sigma_{\rm H2}$, and
$\Sigma_{\rm SFR}$ in a sample of 18 galaxies: 7
spirals which have central areas (hereafter loosely referred to as centers) dominated by molecular gas (`spirals') and 11 \hi-dominated galaxies. We look at how these relationships differ between the H$_2$-dominated centers of the spirals, their \hi-dominated outskirts, and \hi-rich late type galaxies. By probing out to the optical radius $r_{25}$, i.e., where the B-band magnitude drops below 25\,mag\,arcsec$^{-2}$, and including late-type galaxies, we are able
to strongly constrain the universality of the various Schmidt laws.

We organize this analysis as follows: we describe the datasets used
to measure $\Sigma_{\rm HI}$ , $\Sigma_{\rm H2}$, and $\Sigma_{\rm SFR}$
in \S\,\ref{data}. We explain how we
convert these data to physical units and how we
generate a set of independent measurements over the disk of each
galaxy. In \S\,\ref{sflaw-individual} we show the observed relationships between
$\Sigma_{\rm HI}$, $\Sigma_{\rm H2}$, and $\Sigma_{\rm SFR}$
for individual galaxies. We carry out power law
fits to each distribution and assess whether there is a common Schmidt
law shared by all of our galaxies. We show results for spirals and \hi-dominated late-type galaxies separately to highlight the range of
environments in our sample. In \S\,\ref{combined} we combine our data for the spirals and for the \hi-dominated galaxies, to be able to draw general conclusions about \hi , H$_2$, and the SFR in these two samples. In \S\,\ref{environment} we explore the effect of environment by comparing
the Schmidt law in the inner, H$_2$-dominated parts of spirals to
their \hi-dominated outer disks and to \hi-dominated dwarf and
late-type galaxies. We show that the relationship between total gas
and SFR is a clear function of environment and fit the radial
dependence of the molecular-to-atomic gas ratio in spirals. In
\S\,\ref{summary} we summarize our results and give our conclusions. Here we also compare our results to those from previous studies.

\section{Sample, Data, Units, and Measurements}
\label{data}

This work is made possible by the existence of the state-of-the-art multiwavelength data
sets mentioned in the introduction. We use high resolution \hi\ data from `The \hi\
Nearby Galaxy Survey' \citep[\things,][]{WALTER08} and CO maps from
a large ongoing project using HERA on the IRAM~30m telescope \citep[HERACLES,][]{leroy08-2}
and from the BIMA SONG survey \citep[\bima,][]{helfer03}. We trace star formation
combining mid-IR maps from the `{\em Spitzer} Infrared Nearby Galaxies
Survey' \citep[\sings,][]{KENNICUTT03} and FUV maps from the `\galex\
Nearby Galaxy Survey' \citep[NGS,][]{gildepaz07}. In this section we
describe our sample and these datasets, show maps and radial
distributions, and explain how we generate measurements of $\Sigma_{\rm HI}$,
$\Sigma_{\rm H2}$, and $\Sigma_{\rm SFR}$.

\subsection{Sample}
\label{sample-selection}

Our sample of spiral galaxies consists of the cross-section of \things, \sings, the \galex\ NGS
and the HERACLES or BIMA-SONG CO surveys. Galaxies that are known to be \hi-dominated
are not required to be part of HERACLES or BIMA-SONG. This ensures that we measure at least
the dominant component of the neutral ISM and the SFR along most lines
of sight. Table \ref{table-general} lists our sample along with their
adopted properties: distance, inclination, position angle, radius \citep[][except that we adopt $i=20\degr$ in M51]{WALTER08} and morphology \citep[from LEDA,][]{prugniel98}. We separate the galaxies into two groups: 7 large spiral
galaxies that have an H$_2$-dominated ISM in their centers (`spirals') and 11
late-type galaxies that are \hi-dominated throughout.

We do not study galaxies with inclinations $> 70\degr$ that
would otherwise qualify for the sample. High inclinations yield very
few sampling points and result in a deprojected beam elongated parallel to the
minor axis in the plane of the galaxy, making it difficult to carry
out fits to the data and interpret the results. We also do not include any galaxies more distant than
$12$~Mpc. This is the maximum distance at which the typical angular
resolution of our data corresponds to our common
spatial resolution of 750\,pc.

We work without CO maps for 4 late-type spirals, NGC~925, NGC~2403,
NGC~2976, NGC~4214 and 7 dwarf irregular galaxies. For most of these
galaxies, the CO content is either measured or constrained by a
significant upper limit: major axis profiles by \citet{young95} for
NGC~925, NGC~2403, and NGC~2976 (which is also part of BIMA SONG);
Kitt Peak 12m mapping of the inner disk of NGC~2403 \citep{thornley95};
OVRO mapping of NGC~4214 by \citet{walter01}; single dish pointings
toward NGC~4214, Holmberg~I and DDO~154 by \citet{taylor98}; central
pointings for DDO~53, IC~2574, and Holmberg~II by \citet{leroy05}. In
each case, the ISM is well-established to be \hi-dominated. M81~DwA
and M81~DwB, two extremely low-mass dwarf irregular companions to
M81, both lack CO measurements, but should also be \hi-dominated.
Of these galaxies, 8 have absolute
$B$-band magnitudes fainter than $-18$~mag and maximum rotation velocities
$\lesssim 100$~km~s$^{-1}$ (i.e., they have the mass of the LMC or less). NGC~925
and NGC~2403 have $M_{\rm B} \sim -20$~mag and maximum rotational
velocities $\sim 100$~km~s$^{-1}$; they are probably intermediate in
mass between the LMC and M~33.

\begin{deluxetable}{lrrrcc}
\tablecaption{Sample Properties\tablenotemark{1}} \tablehead{
\colhead{Galaxy} & \colhead{$D$} & \colhead{$i$} & \colhead{$PA$} & \colhead{$r_{25}$} & \colhead{Hubble} \\
\colhead{} & \colhead{[Mpc]} & \colhead{[deg]} & \colhead{[deg]} &
\colhead{[arcmin]} & \colhead{type}} \startdata \tableline
\multicolumn{6}{c}{\hi-dominated Galaxies}\\
\tableline
  DDO\,154 & 4.3 & 66 & 230 & 0.97 & Irr \\
  DDO\,53 & 3.6 & 31 & 132 & 0.39 & Irr \\
  Ho\,I & 3.8 & 12 & 50 & 1.66 & Irr \\
  Ho\,II & 3.4 & 41 & 177 & 3.30 & Irr \\
  IC\,2574 & 4.0 & 53 & 56 & 6.44 & SABm \\
  M\,81\,DwA & 3.6 & 23 & 49 & 0.63 & Irr \\
  M\,81\,DwB & 5.3 & 44 & 321 & 0.56 & Irr \\
  NGC\,925 & 9.2 & 66 & 287 & 5.36 & Scd \\
  NGC\,2403 & 3.2 & 63 & 124 & 7.92 & SABc \\
  NGC\,2976 & 3.6 & 65 & 335 & 3.62 & Sc \\
  NGC\,4214 & 2.9 & 44 & 65 & 3.38 & Irr \\
  \tableline
  \multicolumn{6}{c}{Spirals With H$_2$-Dominated Centers}\\
  \tableline
  NGC\,628 & 7.3 & 7 & 20 & 4.89 & Sc \\
  NGC\,3184 & 11.1 & 16 & 179 & 3.71 & SABc \\
  NGC\,3521 & 10.7 & 73 & 340 & 4.16 & SABb \\
  NGC\,4736 & 4.7 & 41 & 296 & 3.88 & Sab \\
  NGC\,5055 & 10.1 & 59 & 102 & 5.87 & Sbc \\
  NGC\,5194 & 8.0 & 20 & 172 & 3.88 & Sbc \\
  NGC\,6946 & 5.9 & 33 & 243 & 5.74 & SABc
\enddata
\tablenotetext{1}{See \citet{WALTER08} for further information on individual galaxies and for references to the values quoted in this table.}
\label{table-general}
\end{deluxetable}

\subsection{Data}
\subsubsection{THINGS \hi\ Maps}
\label{hi}

To estimate the surface density of neutral atomic hydrogen,
$\Sigma_{\rm HI}$, we use VLA maps of the 21\,cm line obtained as part
of `The \hi\ Nearby Galaxy Survey'
\citep[\things,][]{WALTER08}. \things\ consists of high-resolution,
high-sensitivity \hi\ data for 34 nearby galaxies obtained with the
NRAO\footnote{The National Radio Astronomy Observatory is a facility
  of the National Science Foundation operated under cooperative
  agreement by Associated Universities, Inc.} VLA. The target galaxies
have distances of $2\leq\,D\,\leq15$\,Mpc and cover a wide range
in star formation rates, \hi\ masses, luminosities and Hubble
Types.

The FWHM of the primary beam (field-of-view) is 32\,\arcmin . We use
`robust' weighted maps, which have a typical beam size of
$\sim$\,6\,\arcsec. The typical $1\sigma$ RMS noise at our working resolution of 750\,pc (see \S\,\ref{alignment}) is $\sigma(\Sigma_{\rm HI}) \approx 0.5$~M$_{\sun}$~pc$^{-2}$. Because our analysis is restricted to the regime within the optical radius $r_{25}$ (see \S\,\ref{samplingregions} and compare the radial profiles in \S\,\ref{radial-prof}), the measured \hi\ surface densities are safely above this sensitivity limit for all galaxies.

The integrated \hi\ intensity map for NGC~6946 at our working resolution of 750\,pc is shown as an example in the left panel of Figure \ref{fig-1}. Further details regarding
the data products and data reduction are given in \citet{WALTER08}.

\subsubsection{HERACLES CO Maps}
\label{co}

We derive $\Sigma_{\rm H2}$ distributions from two sources: an ongoing
nearby galaxy survey using the HERA focal plane array on the IRAM 30m
telescope \citep[HERACLES,][]{leroy08-2}, and the `BIMA
Survey of Nearby Galaxies' \citep[\bima,][]{helfer03}. The profiles
from the two sets of CO maps agree with one another and also with the
major axis pointings from the FCRAO survey \citep{young95}. We use the
HERACLES maps where available because they have good extent and
sensitivity, often measuring CO well into the \hi-dominated regime
where (on average) $\Sigma_{\rm HI} > \Sigma_{\rm H2}$, i.e., beyond the H$_2$-to-\hi\ transition radius. Extending beyond the transition radius is critical to be able to differentiate between a molecular and total gas Schmidt law.

We use HERACLES maps for 6 spiral galaxies: NGC~628, NGC~3184, NGC~3521,
NGC~4736, NGC~5055 and NGC~6946. These were obtained at the IRAM 30m
telescope during January and October 2007 and further details are described in
\citet{leroy08-2}. We show the CO intensity map for NGC~6946 in the middle panel of Figure \ref{fig-1}. HERA maps the CO~$J=2\rightarrow1$ transition
with an angular resolution of $11\arcsec$. For technical details on
HERA see \citet{schuster2004}. Our maps are sensitive
to surface densities $\Sigma_{\rm H2} \approx
3$~M$_{\odot}$~pc$^{-2}$. 

\bima\ produced maps of CO~$J=1\rightarrow0$ emission with good
resolution, $\sim 7''$, but limited field of view. We use these maps only for
NGC~5194. Also, we use these data to
compare CO to the SFR at high resolution in the central parts of all 7
centrally H$_2$-dominated galaxies. The BIMA SONG map for NGC~5194
used here includes zero-spacing data from the Kitt Peak 12m telescope,
making it sensitive to extended structure. The BIMA
SONG maps we use are sensitive to surface densities above $\Sigma_{\rm H2}
\approx 10$~M$_{\odot}$~pc$^{-2}$.

\subsubsection{GALEX FUV Data}
\label{galex}

We use FUV data from the `\galex\ Nearby Galaxy Survey'
\citep[NGS,][]{gildepaz07} to estimate $\Sigma_{\rm SFR}$ (see
\S\,\ref{sfrmaps}). The \galex\ observatory provides simultaneous
imaging in a far UV (FUV) and a near UV (NUV) broadband filter. The
FUV band covers the wavelength range 1350\,-\,1750\,\AA, the NUV
covers 1750\,-\,2800\,\AA. The angular resolutions (FWHM) are
4.0\,\arcsec\ and 5.6\,\arcsec\ for the FUV and the NUV respectively,
and the field-of-view of the instrument is 1.25\,\degr. For technical
details on the detector see \citet{morrissey05}.

We remove a single set of foreground stars from the GALEX and {\em Spitzer} 24$\mu$m
maps (see \S\,\ref{spitzer}). Foreground stars are prominent in the NUV images but much less
so in the FUV images, and we find that we can identify them easily via
their UV colors. We therefore use the NUV maps to identify foreground star candidates
and apply an NUV/FUV ratio cut of $NUV / FUV > 10$. We apply this to
all emission above 5\,$\sigma$ in the NUV maps, below this we cannot
distinguish stars from noise. In a few cases, a cut-off higher than
10 was necessary to remove particularly bright
foreground stars. We estimate the contribution from foreground
stars below 5\,$\sigma$ in our FUV images beyond
2\,$r_{25}$ (where there is only marginal emission from our target
galaxies) and find that their contribution is negligible.

We estimate and remove a small background from the FUV maps. We
measure this away from the galaxy by discarding emission with
intensities $>3\sigma$ above the median value of the image. The
background adopted is the median of this residual map and is
subsequently subtracted from the original map. The backgrounds in the
GALEX maps are very well behaved and this simple subtraction yields
good results in most cases. For three galaxies -- NGC~4214, NGC~5194
(M51), and NGC~6946 -- we blank portions of the map that show the
edge of the GALEX field-of-view, obvious artifacts, bright stars
not entirely removed by the color-cut as well as M51b. These are
usually located well away from the galaxy and have minimal impact on this study
(which is restricted to the optical disks of the galaxies).

We correct the FUV maps for the effects of Galactic extinction. We estimate
$E(B-V)$ from the extinction maps of \citet{schlegel98} and convert to FUV
extinction using $A_{FUV}=8.24\times E(B-V)$ \citep{wyder07}.

\subsubsection{Spitzer Space Telescope 24 \microm\ Data}
\label{spitzer}

FUV data are heavily affected by internal extinction. To estimate the amount
of star formation obscured by dust (see \S\,\ref{sfrmaps}), we use
24\,\microm\ data obtained by the `{\em Spitzer} Infrared Nearby Galaxies
Survey' \citep[SINGS,][]{KENNICUTT03}. The map of NGC~4214 comes from the public
archive. These data are scan maps taken with the MIPS instrument
on board the \spitzer\ Space Telescope \citep{rieke04}. Full details of the observing strategy
and target sensitivity are described by \citet{KENNICUTT03}. The MIPS data
were processed using the MIPS Instrument Team Data Analysis Tool
\citep{gordon05}.

The \sings\ observations were designed to detect emission out to the
optical radius and in most of our targets this goal was achieved. The
FWHM of the MIPS PSF at 24\,\microm\ is 6\,\arcsec, though the beam
is substantially non-Gaussian outside the central peak. This is only a
minor concern for this study because to achieve a common spatial
resolution for all data, we work with a typical angular resolution
of $\sim 20''$.

We perform the following additional processing on the SINGS maps: we first
blank the edges of the scan maps parallel to the direction of the scan; in
these regions the noise increases and artifacts are more common. We also blank
the stars identified from the GALEX images. We then subtract a background, which
is usually quite small, using the same procedure as for the GALEX images. In a
few maps -- Ho~I, NGC~3521 and NGC~6946 -- we identify regions with unreliable
background and blank these by hand. These regions are far away from
the galaxy though and so have minimal effect on this analysis.

\subsection{Alignment, Units, and Convolution}
\label{alignment}

From the data described above, we construct maps of star formation
rate surface density, $\Sigma_{\rm SFR}$, atomic hydrogen surface
density, $\Sigma_{\rm HI}$, molecular hydrogen surface density,
$\Sigma_{\rm H2}$, and total gas surface density, $\Sigma_{\rm gas}$.

We place all of these maps on the \things\ astrometric grid (pixel
scale: $1.5\,\arcsec$) at a common spatial resolution of
750~pc. We carry out most of our work at 750~pc resolution but also
create versions of each map at a range of spatial resolutions, from the
native resolution to 10~kpc (in steps of 50~pc below 1~kpc, and 500~pc
up to 10~kpc). When degrading the resolution of a map, we convolve
with a circular Gaussian beam (on the sky), i.e., we do not account for the
inclination of the galaxy. For a given spatial resolution, this
exercise may be thought of as moving each galaxy at a larger distance.

\subsubsection{Gas Surface Density Maps}

The total gas surface density, $\Sigma_{\rm gas}$, is the sum of the
atomic gas surface density, $\Sigma_{\rm HI}$, and the molecular gas
surface density, $\Sigma_{\rm H2}$ (we do not consider ionized gas, as this
makes up only a small fraction of the total gas content).
Where $\Sigma_{\rm H2}$ is below the sensitivity limit (see \S\,\ref{co}),
we take $\Sigma_{\rm gas}=\Sigma_{\rm HI}$, which is formally a lower limit. 
These quantities all have units of M$_{\odot}$~pc$^{-2}$ and are
{\em hydrogen} surface densities, i.e., they do not include any
contribution from helium. To scale our quoted surface densities to
account for helium, one should multiply them by a factor of $\sim
1.36$. All surface and column densities quoted in this paper have been
corrected for the inclinations given in Table \ref{table-general}.

We assume a ratio $I(2\rightarrow1) / I(1\rightarrow0) = 0.8$, a typical number derived from comparing our maps to the maps of \citet{helfer03}, \citet{young95} and \citet{kuno07}. A detailed comparison between these datasets can be found in \citet{leroy08-2}. To convert from $\Sigma_{\rm CO}$ to $\Sigma_{\rm H2}$, we adopt a CO-to-H$_2$ conversion factor of $2.0 \times 10^{20}$~\xcounits. This
value is appropriate for the Milky Way according to $\gamma$-ray and
FIR studies \citep[e.g.,][]{strong96,dame01}, but slightly lower than
the conversion factor suggested by virial mass methods
\citep[e.g.,][]{solomon87,blitz07}. We do not account for any changes
in the conversion factor with metallicity or other environmental
factors. For comparison, we note that we use the same conversion
factor adopted by \citet{wong02} but that K98 adopted a higher value
of $2.8 \times 10^{20}$~\xcounits.

\subsubsection{Star Formation Rate Surface Density Maps}
\label{sfrmaps}

We measure star formation rate surface densities, $\Sigma_{\rm SFR}$,
in units of M$_{\odot}$~yr$^{-1}$~kpc$^{-2}$, by combining FUV and
24$\mu$m maps. The \galex\ FUV images trace predominantly O and early
B stars and therefore offer a picture of recent, unobscured star
formation. FUV emission can be heavily affected by dust, however,
making it difficult to derive unbiased measurements of $\Sigma_{\rm
  SFR}$ from the FUV alone. Fortunately, the dust that obscures
the FUV emission from young stars is heated and reradiates in the
mid-infrared. The 24\,\microm\ maps thus allow us to estimate the
amount of ongoing dust-obscured star formation.

This approach was proposed and validated for individual star forming
regions and large portions of galactic disks by \citet{CALZETTI07} and
\citet{KENNICUTT07}. They showed that 24\,\microm\ emission could be
used to accurately estimate the amount of extinction that affects
H$\alpha$ emission.  We adopt an analogous approach and combine FUV
and 24\,\microm\ using the following formula:

\begin{eqnarray}
\label{eq-sfr}
\Sigma_{\rm SFR} \left[ {\rm M}_\odot~{\rm yr}^{-1}~{\rm kpc}^{-2}\right]
&=&3.2 \times 10^{-3} I_{24} \left[{\rm MJy~ster}^{-1}\right] \\
\nonumber &+& 8.1 \times 10^{-2} I_{FUV} \left[{\rm MJy~ster}^{-1}\right]~,
\end{eqnarray}

\noindent where $\Sigma_{\rm SFR}$ denotes the star formation rate
surface density, and I$_{24}$ and I$_{FUV}$ are the 24~\microm\ and
FUV intensities respectively. \citet{leroy08} motivate the choice of
coefficients in Equation \ref{eq-sfr}. $\Sigma_{\rm SFR}$ derived from
Equation \ref{eq-sfr} agrees with the values derived using the
\halpha\,+\,24\,\microm\ calibration from \citet{CALZETTI07} at 750~pc
resolution; when $I_{24} = 0$, Equation \ref{eq-sfr} reduces to the
FUV-SFR calibration by \citet{salim07}. SFRs integrated over entire
galaxies as well as values derived from azimuthally averaged radial profiles
agree well with those based on extinction-corrected \halpha\ emission alone.

As an example we show the $\Sigma_{\rm SFR}$ map for NGC~6946, which was derived using Equation
\ref{eq-sfr}, in the right panel of Figure
\ref{fig-1}. The RMS noise varies from map to map, but a typical limit
for the sensitivity of the SFR maps is $\approx 1 \times 10^{-4}$
\,M$_{\sun}$\,yr$^{-1}$\,kpc$^{-2}$. For our calibration of $\Sigma_{\rm SFR}$
we adopt the IMF from \citet{CALZETTI07}, which is the default IMF in STARBURST99 \citep{leitherer99},
i.e., a Kroupa-type two-component IMF that extends to 120\,M$_\odot$. To convert to the truncated
\citet{salpeter55} IMF adopted by, e.g.,
\citet{KENNICUTT89,KENNICUTT98} or \citet{KENNICUTT07}, one should
multiply our $\Sigma_{\rm SFR}$ by a factor of 1.59.

Because this particular combination of FUV and mid-IR maps is new, we
compare our results to those obtained using maps of H$\alpha$ and
H$\alpha$+24$\mu$m emission wherever these are available. For this purpose, we use H$\alpha$ maps
from the SINGS Data Release 4. We fit and remove
backgrounds from these maps, masking out any region with a
particularly problematic background. We correct for [NII] in the
bandpass following \citet{CALZETTI07} and for Galactic extinction using
the dust maps of \citet{schlegel98}. We use H$\alpha$ maps for all of
our spiral galaxies except NGC~4736.

In addition to sampling our data on a pixel-by-pixel basis at 750~pc
resolution, we also show the results of our analysis carried out in
azimuthally-averaged radial profiles and explore the effects of
degrading to lower resolution (so that each `pixel' contains many
square kpc). Further, we compare our results to measurements from other
studies using a variety of SF tracers (see \S\,\ref{discussion-previous}).
These tests are shown in the appropriate
sections below and lead us to conclude that our SFR maps are
accurately tracing the amount and distribution of recent star
formation.

\subsection{Individual Data Points}
\label{samplingregions}

\begin{figure*}
\epsscale{1.15}
\plotone{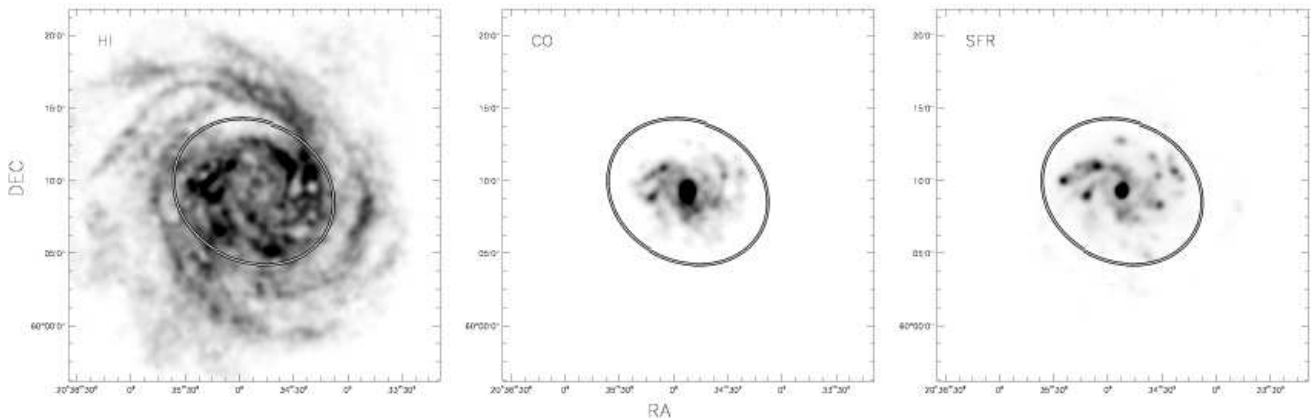}
\caption{Maps of integrated \hi\ (left), CO (middle) and SFR (right) surface densities
  at our working resolution of 750\,pc for the spiral galaxy NGC~6946. The ellipses indicate
  the optical radius ($r_{25}$) in the plane of the galaxy, within which we compare $\Sigma_{\rm HI}$,
  $\Sigma_{\rm H2}$, and $\Sigma_{\rm SFR}$. Almost all star formation occurs within this radius,
  although the \hi\ often extends much beyond $r_{25}$. The maps for all other galaxies in our
  study are shown in \citet{leroy08}.}
\label{fig-1}
\end{figure*}

With maps of $\Sigma_{\rm HI}$, $\Sigma_{\rm H2}$, and $\Sigma_{\rm
  SFR}$ in hand, the next step is to
generate individual data points. We compare these quantities over the
entire optical disk of each galaxy, out to $r_{25} = d_{25}/2$. This
is a departure from previous studies, which used integrated
measurements over entire galaxy disks
\citep[e.g.,][]{KENNICUTT89,KENNICUTT98}, radial profiles
\citep[e.g.,][]{wong02}, or apertures centered on \halpha\ and 24\,\microm\ emission peaks
\citep[e.g.,][]{KENNICUTT07}.

We choose $r_{25}$ as an outer limit. Most star formation takes place within
this radius \citep[e.g., see the profiles of][]{martin01,wong02} and our FUV,
24$\mu$m, and \hi\ maps detect emission at more than 3$\sigma$ over most of this area.
The ellipses in Figure \ref{fig-1} show the projected
optical radius, i.e., the extent of the region that we study, on the \things\
\hi, HERACLES CO and $\Sigma_{\rm SFR}$ maps for NGC~6946. We refer the reader to
\citet{leroy08} for a `galaxy atlas' showing all $\Sigma_{\rm HI}$, $\Sigma_{\rm H2}$, and
$\Sigma_{\rm SFR}$ maps for the galaxies in our sample.

We draw independent data points from our maps so that each data point
corresponds to non-overlapping resolution elements and the data together cover
the optical disk. This may be thought of as either reducing the oversampled
map (rebinning the map) so that one pixel corresponds to a resolution element or
as covering the disk with non-overlapping apertures equal in size to a
resolution element. For each independent data point we measure $\Sigma_{\rm
  HI}$, $\Sigma_{\rm H2}$, $\Sigma_{\rm gas} = \Sigma_{\rm HI} + \Sigma_{\rm
  H2}$, and $\Sigma_{\rm SFR}$.

When convolving the maps with progressively larger beams (see
\S\,\ref{alignment}), flux from outside our radius cut at
$r_{25}$ would be convolved into the optical disk. To avoid this,
the radius cut we use is decreased by half a beam width and
is actually $r_{max} = r_{25} - 0.5\,\theta_{\rm beam}$, where
$\theta_{beam}$ is the FWHM of the beam in the same units as the
radius. Note that $\theta_{\rm beam}$ is subtracted in the plane of
the sky, not the plane of the galaxy because the convolution takes
place in the plane of the sky.

\subsection{Radial Profiles}
\label{radial-prof}

\begin{figure*}
  \plotone{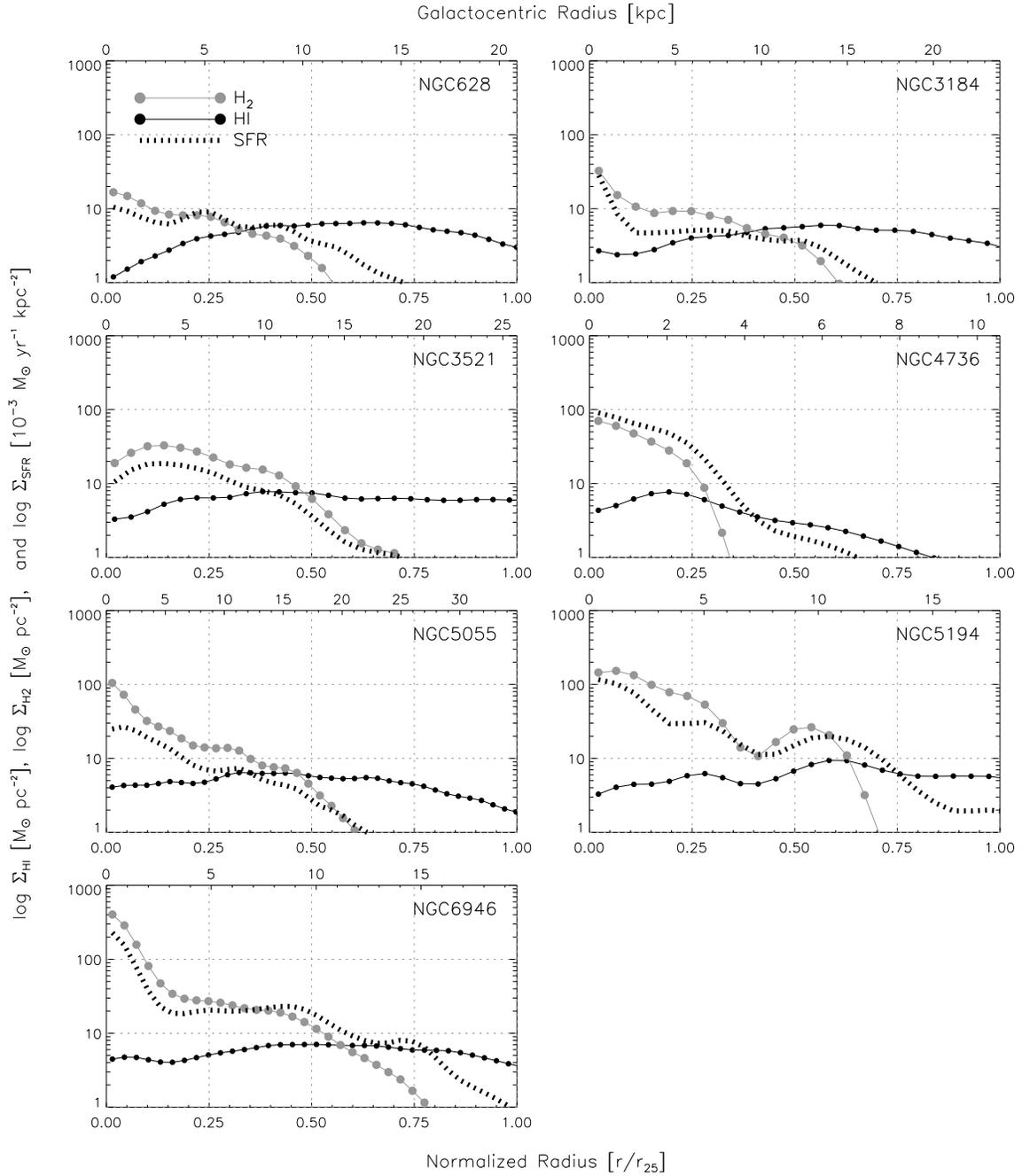}
  \caption{Azimuthally averaged radial profiles of $\Sigma_{\rm HI}$,
    $\Sigma_{\rm H2}$, and $\Sigma_{\rm SFR}$ for spiral galaxies with
    H$_2$-dominated centers. The $y$-axis shows $\Sigma_{\rm HI}$ and
    $\Sigma_{\rm H2}$ in units of M$_{\odot}$~pc$^{-2}$ as well as $\Sigma_{\rm SFR}$
    in units of 10$^{-3}$~M$_{\odot}$~yr$^{-1}$~kpc$^{-2}$ (the scaling is
    chosen to bring the profiles onto the same plot). The $x$-axis
    shows galactocentric radius normalized by $r_{25}$ (bottom) and
    in kpc (top). Profiles of $\Sigma_{\rm SFR}$ and
    $\Sigma_{\rm H2}$ are strongly covariant. $\Sigma_{\rm HI}$ varies
    weakly over the optical disk with the main features often being a
    central depression and a universal upper limit of $\Sigma_{\rm HI}
    \approx 9$~M$_{\odot}$~pc$^{-2}$.}
\label{fig-11}
\end{figure*}

\begin{figure*}
\plotone{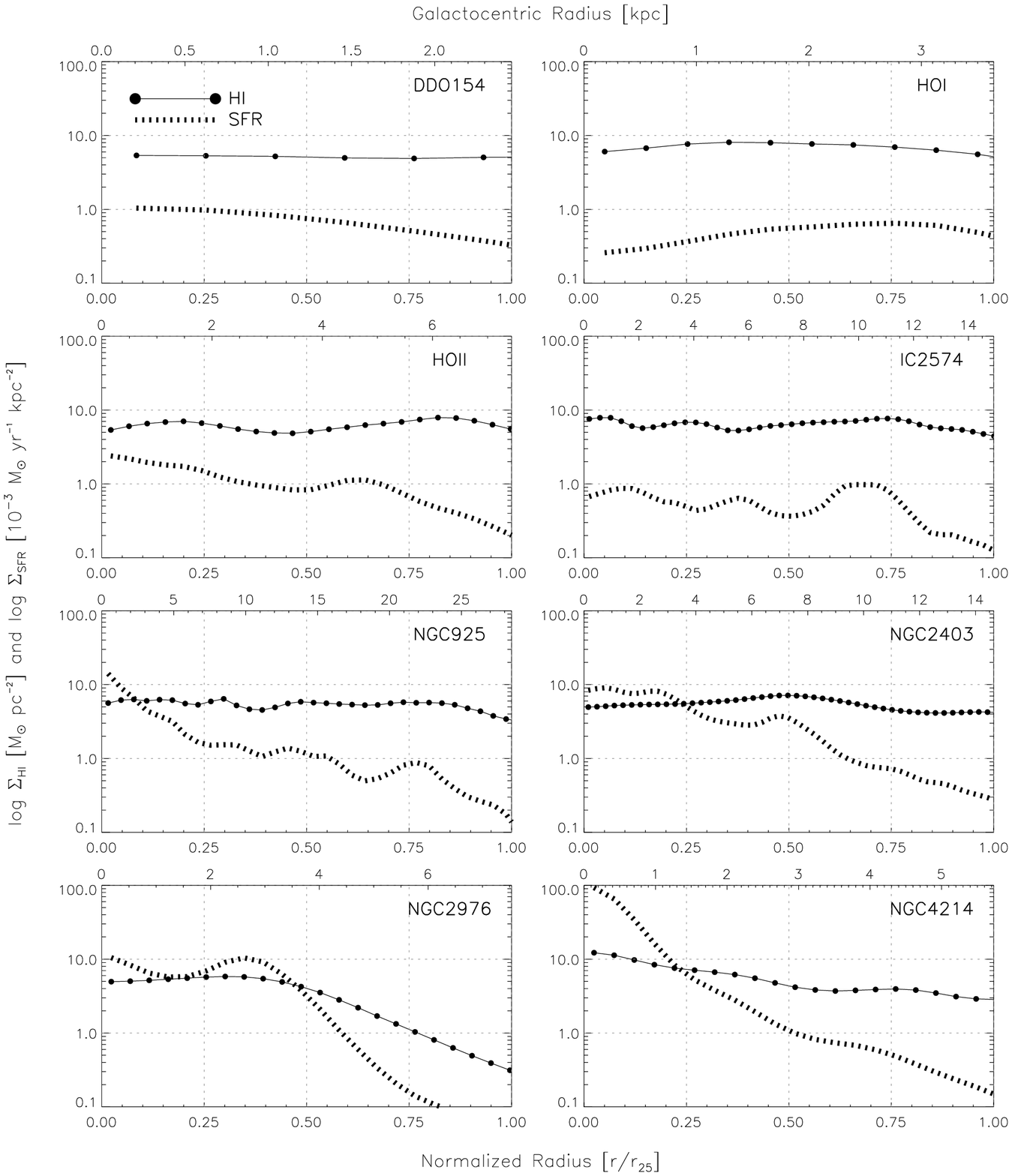}
\caption{Azimuthally averaged radial profiles of $\Sigma_{\rm HI}$ and
  $\Sigma_{\rm SFR}$ for 8 of our 11 \hi-dominated galaxies, all
  late-type spirals or dwarf irregulars. The $y$-axis shows $\Sigma_{\rm HI}$
  in units of M$_{\odot}$~pc$^{-2}$ or $\Sigma_{\rm SFR}$
  in units of 10$^{-3}$~M$_{\odot}$~yr$^{-1}$~kpc$^{-2}$ (the scaling is
  chosen to bring the profiles onto the same plot). The $x$-axis shows
  the galactocentric radius normalized by $r_{25}$ (bottom) and in
  kpc (top). $\Sigma_{\rm HI}$ shows the same maximum value seen in
  spiral galaxies and similarly flat profiles.}
\label{fig-12}
\end{figure*}

We also extract from our maps radial profiles of $\Sigma_{\rm HI}$,
$\Sigma_{\rm H2}$, and $\Sigma_{\rm SFR}$ and examine them along with
the pixel-by-pixel data. We show the profiles for the spirals and 8 of our \hi-dominated
galaxies in Figures \ref{fig-11} and \ref{fig-12}. We note that the radial profiles as well as the \hi, CO and SFR maps for all of our galaxies are shown in \citet{leroy08-2}. The plots in Figures \ref{fig-11} and \ref{fig-12} show
$\Sigma_{\rm SFR}$, $\Sigma_{\rm HI}$ and $\Sigma_{\rm H2}$ as a
function of galactocentric radius normalized by $r_{25}$. The top axis gives the radius in
kpc. Each point represents the average value in an individual
$10\arcsec$-wide tilted ring within 60\degr of the major axis using the structure parameters given in Table \ref{table-general}. For the \hi-dominated galaxies, we show only
$\Sigma_{\rm SFR}$ and $\Sigma_{\rm HI}$.

\section{The Star Formation Law in Individual Galaxies}
\label{sflaw-individual}


Figure \ref{fig-2} shows the relationship between gas and star
formation surface densities in individual spiral galaxies. Color
contours show the results from pixel-by-pixel sampling the optical
disks of the 7 spiral galaxies in our sample. Each row shows results
for a different galaxy. The columns left to right show $\Sigma_{\rm
  SFR}$ on the y-axis as a function of $\Sigma_{\rm HI}$ (left), $\Sigma_{\rm H2}$
(middle), and $\Sigma_{\rm gas}$ (right) on the x-axis. Shading shows the density of
data points in cells of size 0.05~dex wide (in both axes).  Green,
orange, red, and magenta cells show contours of 1, 2, 5, and 10
sampling points per cell.  Individual cells that are not connected and
contain only one data point are marked in grey. All data in Figure
\ref{fig-2} are at a common resolution of 750\,pc.

We plot points from the radial profiles (Figures \ref{fig-11} and
\ref{fig-12}) on the same plot as black crosses. Generally speaking,
the radial profile data lie near the middle of the distribution of
the pixel-by-pixel data, as expected.  In many cases, features at high
$\Sigma_{\rm gas}$ and high $\Sigma_{\rm SFR}$ are clearer in the
radial profile data because small central rings with high surface
densities have small areas and thus contribute only a few individual
points to the pixel-by-pixel data.

A common way to parametrize the relationship between star formation
and gas is via the gas depletion time or its inverse, the star
formation efficiency ${\rm SFE} = \Sigma_{\rm SFR} / \Sigma_{\rm gas}$. The gas depletion time
is the time needed for the present rate star formation to consume the existing
gas reservoir. In Figure \ref{fig-2}, we plot levels of constant
depletion time/SFE as diagonal dotted lines. From bottom to top, these correspond to gas depletion
times of 10$^{10}$, 10$^{9}$ and 10$^{8}$\,years or
equivalently depleting 1\%, 10\% and 100\% of the gas reservoir per
10$^{8}$\,years; these values include helium and so are true
depletion times.

\begin{figure*}
\plotone{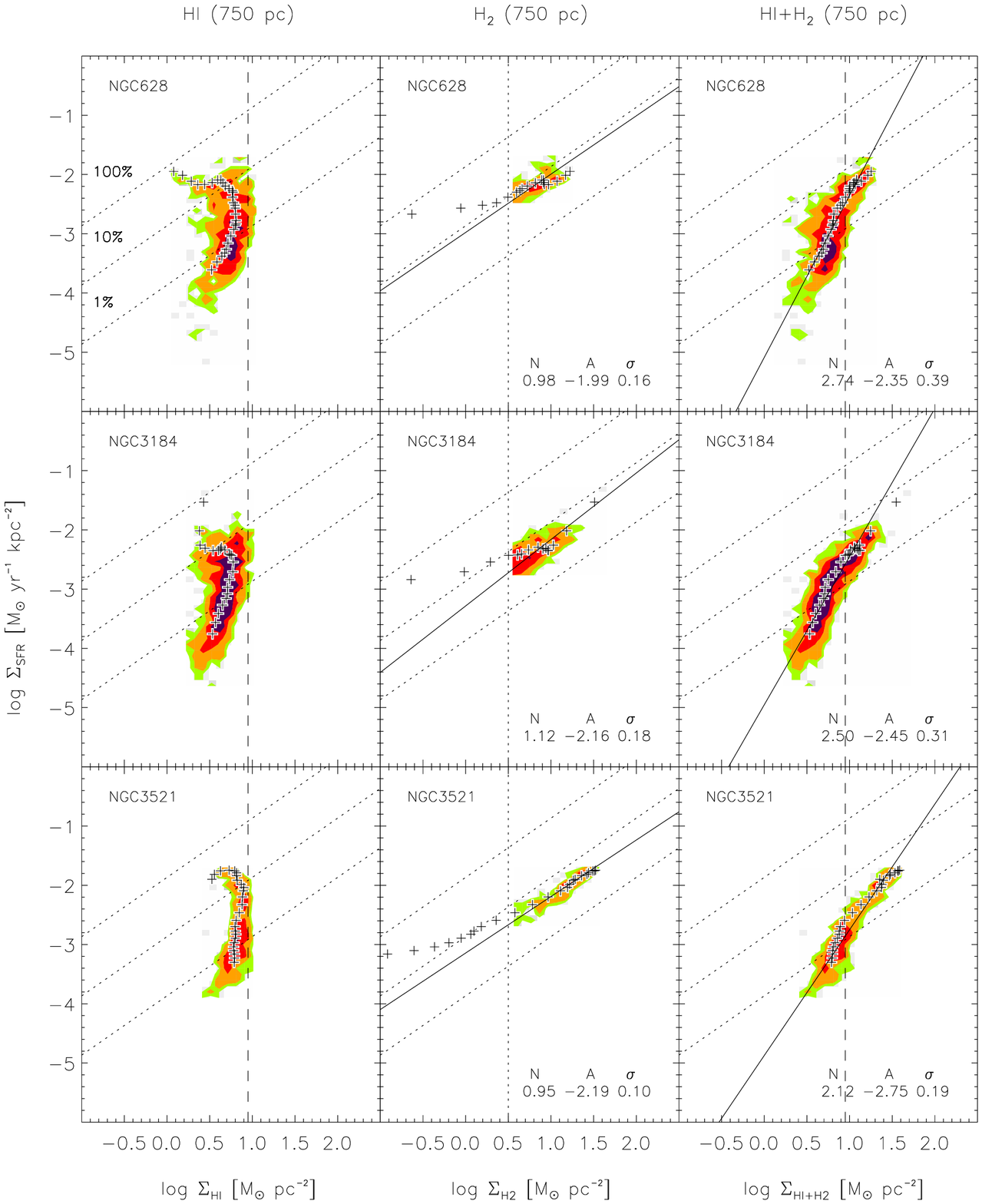}
\caption{$\Sigma_{\rm SFR}$ as a function of $\Sigma_{\rm HI}$ (left),
  $\Sigma_{\rm H2}$ (middle), and $\Sigma_{\rm gas} = \Sigma_{\rm HI}
  + \Sigma_{\rm H2}$ (right) in our spiral galaxies at 750~pc
  resolution. Each row shows results for one galaxy.  Green, orange,
  red, and magenta cells show contours of 1, 2, 5, and 10 independent
  data points per $0.05$~dex-wide cell (for H$_2$ in NGC~4736 we use a
  scatter plot due to the low number of sampling points.). Crosses
  show average measurements over tilted rings from the radial
  profiles. Diagonal dotted lines show lines of constant SFE, indicating the level of
   $\Sigma_{\rm SFR}$ needed to consume 1\%, 10\% and 100\% of the gas reservoir
  (including helium) in 10$^{8}$\,years. Thus, the lines also correspond to constant
gas depletion times of, from top to bottom, $10^{8}$, $10^{9}$, and $10^{10}$~yr. Dashed vertical lines in the \hi\ (left) and total gas
  (right) plots show the surface density where the \hi\
  saturates (see \S\,\ref{saturation}). Dotted vertical lines in the middle plots show the typical
  sensitivity for our CO data.  We show OLS bisector fits to the
  \htwo\ and total gas data with a solid line and quote the
  results.}
\label{fig-2}
\end{figure*}

\begin{figure*}
\figurenum{\ref{fig-2}}
\plotone{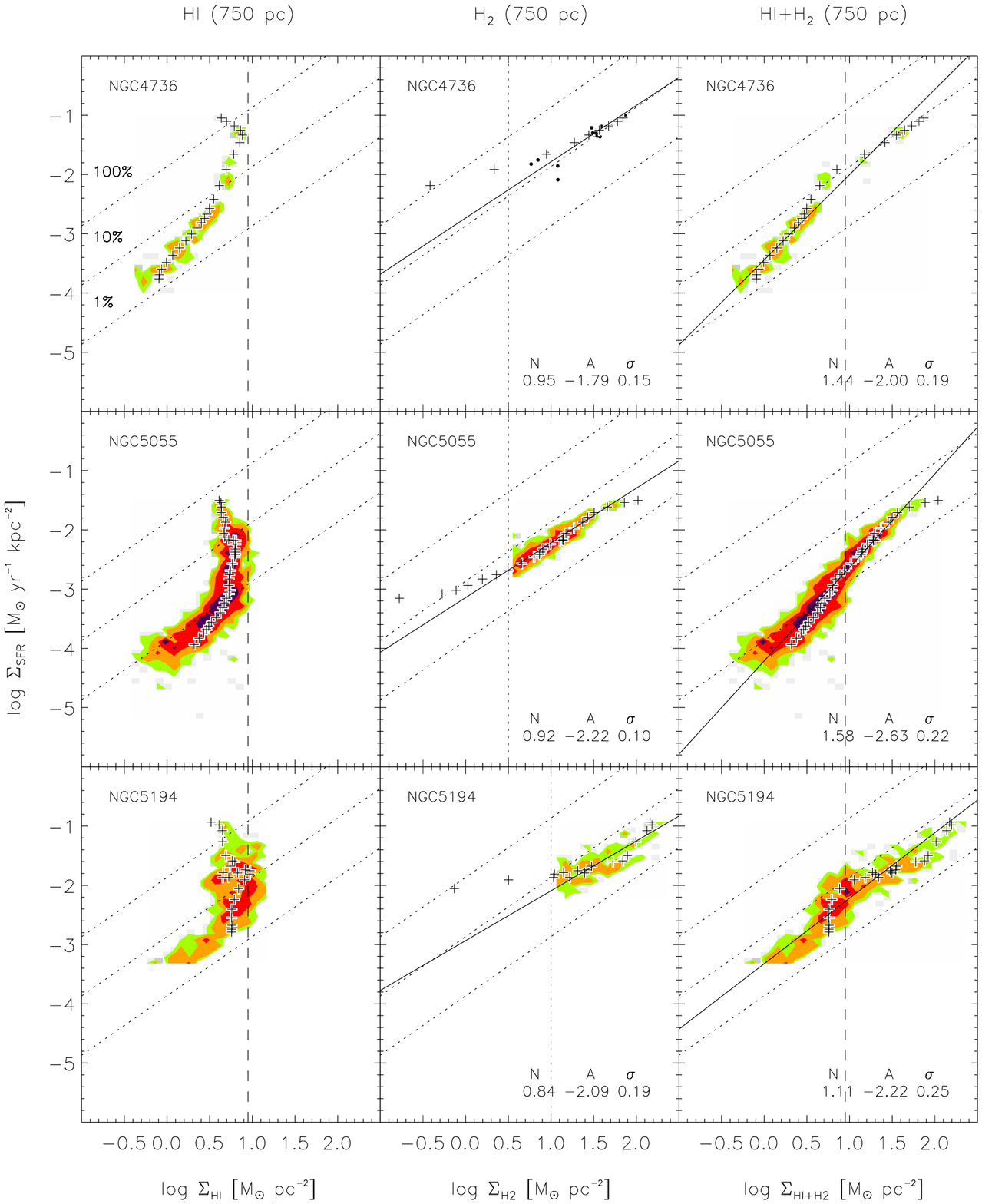}
\caption{continued.}
\end{figure*}

\begin{figure*}
\figurenum{\ref{fig-2}}
\plotone{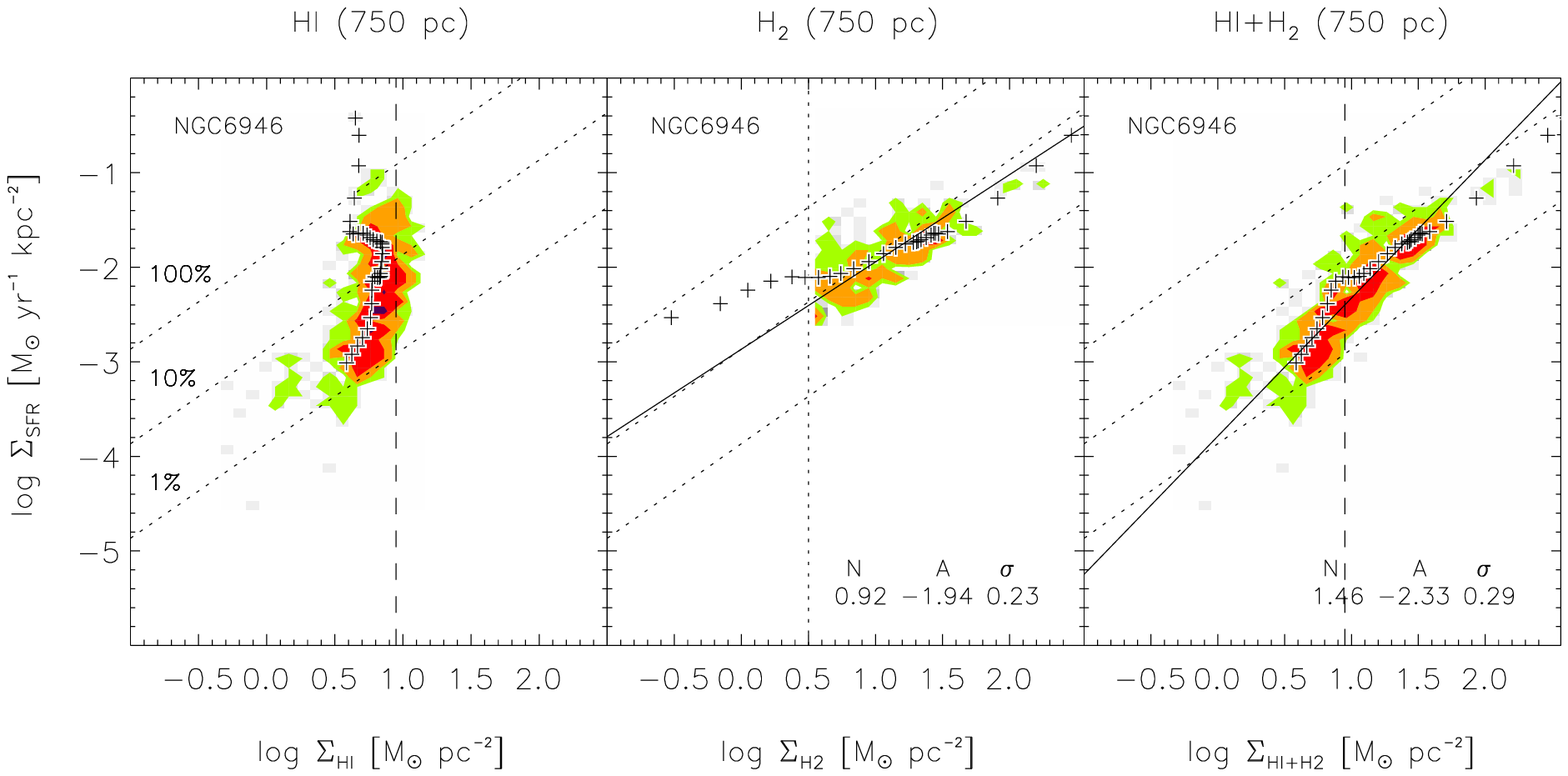}
\caption{continued.}
\end{figure*}

\subsection{Fits to the $\Sigma_{\rm SFR}$ Versus $\Sigma_{\rm gas}$ Distributions}
\label{fits-individual}

Solid black lines in Figure \ref{fig-2} show the results of fitting a
power-law of the form

\begin{equation}
\label{power-law}
\Sigma_{\rm  SFR} = a~\left( \frac{\Sigma_{\rm HI,H2,gas}}{10~{\rm M}_{\odot}~{\rm pc}^{-2}} \right)^{N}~.
\end{equation}

\noindent using the ordinary least--squares (OLS) bisector. The free
parameters are the power law index, $N$, and $a$, which is
$\Sigma_{\rm SFR}$ at the fiducial gas surface density of
$10$~M$_{\odot}$~pc$^{-2}$.  Figure \ref{fig-2} and Table
\ref{table-fits} give the best fit values of $N$, $A= \log_{10} a$,
and the RMS scatter in $\log \Sigma_{\rm SFR}$ about the fit. Because there is not a clear independent variable, we carry out the fit using
the OLS bisector \citep{isobe90}, giving equal weight to each point. We treat
the problem in log space, where fitting a line yields $N$ as the slope and $A$
as the intercept. By centering the fit at
$10$~M$_{\odot}$~pc$^{-2}$, a surface density near the middle of the distributions, we
minimize the covariance between $N$ and $a$. We note that readers interested in
comparison with previous work should take note of this difference; most fits
in the literature quote `$A$' at $1$~M$_{\odot}$~pc$^{-2}$ rather than our
fiducial $10$~M$_{\odot}$~pc$^{-2}$.

We fit $\Sigma_{\rm SFR}$ vs. $\Sigma_{\rm H2}$ (middle panel) in the
regime $\Sigma_{\rm H2} > 3$~M$_{\sun}$\,pc$^{-2}$ for the HERACLES data and
$\Sigma_{\rm H2} > 10$~M$_{\sun}$\,pc$^{-2}$ for the \bima\ data (see \S\,\ref{co}). These sensitivities are shown as a dotted vertical line in the middle panels of
Figure \ref{fig-2}. We fit $\Sigma_{\rm SFR}$ vs. $\Sigma_{\rm gas}$
to all data points (right panel). The formal errors on the fits are small and the fits are
robust to the removal of individual data points. We test the latter
via bootstrapping (i.e., repeatedly drawing a new, equal-sized, random subsample from our data allowing repetition) and find the resulting uncertainties to be typically
$\sim 0.01$ in $N$ and $\sim 0.05$ in $A$.  Methodology, e.g., the
decision to fit X vs. Y, Y vs. X, or use of the OLS bisector, drives the
resulting fits as much as any other statistical factor.  Because both
variables are independent, we use the OLS bisector, but estimate the
uncertainty by carrying out ordinary least-squares (OLS) fits of
$\Sigma_{\rm SFR}$ vs.  $\Sigma_{\rm H}$ and $\Sigma_{\rm H}$
vs. $\Sigma_{\rm SFR}$. The differences in $N$ and $A$ that we obtain
are taken as the uncertainties on the fit. We find that this brackets the
range of reasonable `by eye' fits well. We note that these uncertainties represent
only the uncertainty in fitting the distribution of data points. They do not
reflect systematics such as uncertainty or variations in the
CO-to-H$_2$ conversion factor, the IMF, etc.

\begin{deluxetable*}{lcccccc}
\tablecaption{Fitted Power-Law Parameters at 750\,pc Resolution}
\tablehead{
\colhead{} & \multicolumn{3}{c}{H$_{2}$} & \multicolumn{3}{c}{\hi\,+\,H$_{2}$} \\
\colhead{Galaxy} & \colhead{Coefficient ($A$)} & \colhead{Index ($N$)} & \colhead{Scatter} & \colhead{Coefficient ($A$)} & \colhead{Index ($N$)} & \colhead{Scatter}}
\startdata
NGC\,628 & $-1.99$ & $0.98$ & $0.16$ & $-2.35$ & $2.74$ & $0.39$ \\
NGC\,3184 & $-2.16$ & $1.12$ & $0.18$ & $-2.45$ & $2.50$ & $0.31$ \\
NGC\,3521 & $-2.19$ & $0.95$ & $0.10$ & $-2.75$ & $2.12$ & $0.19$ \\
NGC\,4736 & $-1.79$ & $0.95$ & $0.15$ & $-2.00$ & $1.44$ & $0.19$ \\
NGC\,5055 & $-2.22$ & $0.92$ & $0.10$ & $-2.63$ & $1.58$ & $0.22$ \\
NGC\,5194 & $-2.09$ & $0.84$ & $0.19$ & $-2.22$ & $1.11$ & $0.25$ \\
NGC\,6946 & $-1.94$ & $0.92$ & $0.23$ & $-2.33$ & $1.46$ & $0.29$ \\
\hline\\
Average & $-2.06 \pm 0.17$ & $0.96 \pm 0.07$ & \nodata & $-2.39 \pm 0.28$ & $1.85 \pm
0.70$ & \nodata
\enddata
\label{table-fits}
\end{deluxetable*}

\subsection{The Molecular Gas Schmidt Law}
\label{mol-gas-schmidt-indiv}

Fitting $\Sigma_{\rm SFR}$ to $\Sigma_{\rm H2}$ alone yields power law
indices near unity, $N=0.96 \pm 0.07$ and coefficients $A = -2.06 \pm
0.17$ with a typical scatter of $\sim 0.2$~dex. That is our 7 spirals
display power law indices consistent with an $N=1$ molecular Schmidt law and only mild
variations in the normalization. Another statement
of this is that the molecular gas in our sample shows a nearly
constant ratio of $\Sigma_{\rm SFR}$ to $\Sigma_{\rm H2}$ that
corresponds to star formation consuming the H$_{2}$ gas reservoir in $\sim
2 \times 10^9$~yrs (i.e., a constant gas depletion time; also see \S\,\ref{sflaw-molecular}).

\subsection{The Total Gas Schmidt Law}
\label{total-gas-individual}

The relationship between $\Sigma_{\rm gas}$ and $\Sigma_{\rm SFR}$
shows a much larger range of behavior as in the H$_{2}$ case. The
power law indices, $N$, for these fits range from 1.11 to 2.74 with the
mean $1.85 \pm 0.70$ and a scatter of $\sim 0.3$~dex. This range is similar
to that found by studies of individual galaxies in the literature
($1\lesssim N\lesssim3$, see \S\,\ref{intro}). These steeper fits, as compared to the H$_{2}$ case, are mainly caused by a drop in $\Sigma_{\rm SFR}$ over a relatively narrow
range of gas surface densities just below $\Sigma_{\rm gas} \sim
9$~M$_{\odot}$~pc$^{-2}$ (see \S\,\ref{saturation} where we assess this further ).

There is less variation in the coefficient, $A$, than in the power law
index. $A$ varies from $-2.75$ to $-2.00$ with a mean of $-2.39$ and an RMS scatter of 0.28. This mean value is lower than the mean value for the molecular gas $A=-2.06$ (see \S\,\ref{mol-gas-schmidt-indiv}). The ratio of mean coefficients $a$ for the total gas and the H$_2$
is $\sim 0.3$ dex; that is, a particular surface density of molecular
gas will on average form stars at about twice the rate of the same surface
density of total gas.

It is further evident that there is no universal behavior in the right
hand column of Figure \ref{fig-2}. That is, the distribution of points
in $\Sigma_{\rm gas}$-$\Sigma_{\rm SFR}$ parameter space varies from
galaxy to galaxy. In some cases, e.g., NGC~5194, a single power law
appears to relate the two. In other cases, e.g., NGC~3184,
$\Sigma_{\rm SFR}$ and $\Sigma_{\rm gas}$ are essentially uncorrelated where
$\Sigma_{\rm HI} > \Sigma_{\rm H2}$. In these cases, there is
no clear one-to-one relationship between total gas and SFR across
the whole disk.

What causes some galaxies to display a power law spanning from the \hi-dominated
to H$_2$-dominated ISM while others do not? The data-driven
answer is that galaxies which show a single power law have
comparatively low \hi\ surface densities within their optical
disks. Galaxies with uncorrelated $\Sigma_{\rm gas}$-$\Sigma_{\rm SFR}$
distributions (e.g., NGC~3184) by contrast do not have many lines of sight with
\hi\ at low surface densities within their optical disks.

One possible explanation is that galaxies with well-defined total gas Schmidt laws (and low \hi\
surface densities) may have lost diffuse \hi\ unassociated with star
formation in interactions. Many show signs of tidal disruption or
ongoing interactions. Assessing the underlying reason for the range of
distributions in the right hand part of Figure \ref{fig-2} is beyond
the scope of this paper, the key observation here is that there {\em
  is} a range of distributions and that many galaxies are not well
described by a single power law relating $\Sigma_{\rm gas}$ to $\Sigma_{\rm SFR}$.

\subsection{Star Formation Efficiencies}
\label{sfe}

The SFE provides another way to express the results discussed
above. Most galaxies show a fixed SFE relating their $\Sigma_{\rm SFR}$ and
$\Sigma_{\rm H2}$. Some galaxies (e.g., NGC~5194) also show a constant SFE in their
total gas spanning from high to low surface densities. The case of
NGC~5194 is particularly striking; this galaxy displays a nearly
constant SFE spanning surface densities from $1$~M$_{\odot}$~pc$^{-2}$
to $100$~M$_{\odot}$~pc$^{-2}$. Over
two orders of magnitude in gas surface density, the power law index
remains near unity and $\Sigma_{\rm SFR}$ shows only a factor of $\sim
2$ scatter about a constant SFE.

On the other hand, several galaxies, e.g., NGC~628, NGC~3184 or NGC~3521,
show steep distributions in the right hand column of
Figure \ref{fig-2} and high power law indices. This may be phrased as
large internal variations in their SFE. The variations in SFE are as
striking in their own way as the correlation in NGC~5194: these
galaxies span nearly an order of magnitude in SFE at an almost
constant $\Sigma_{\rm gas} \approx 5$~M$_\odot$~pc$^{-2}$. This is clear
evidence that the total gas surface density {\em cannot} be the
critical quantity setting the SFR over the \hi-dominated parts of
these galaxies.

There is also variation in the SFE {\em among} galaxies. This can be seen
from the range of coefficients to our power law fits. At a particular
$\Sigma_{\rm gas}$, the average $\Sigma_{\rm SFR}$ shows an RMS scatter of
$\sim 0.3$ dex. Galaxy-to-galaxy variations thus account for a factor of $\sim 2$
scatter in the SFE in our sample.

\subsection{\hi\ Saturation at High Column Densities}
\label{saturation}

Figure \ref{fig-2} and the radial profiles in Figure \ref{fig-11} also
illuminate the relationship between $\Sigma_{\rm HI}$ and $\Sigma_{\rm
  H2}$.  Both show a striking absence of high surface density \hi ;
this is seen as a sharp right-hand edge to the distributions shown
in the left-hand column of Figure \ref{fig-2} and the failure of
our radial profiles (spirals or \hi-dominated galaxies) to
cross $\Sigma_{\rm HI} \approx 9$~M$_{\odot}$~pc$^{-2}$. The only gas in
excess of this limiting surface density appears to be in the molecular
phase. \citet{wong02} showed a similar `saturation' effect in
azimuthally averaged profiles for their molecular gas-rich spirals, as did, e.g.,
\citet{martin01} and \citet{morris78}. The data plotted in Figure \ref{fig-2} show that this
effect is present at $750$~pc resolution and that it is remarkably
universal. $\Sigma_{\rm SFR}$ and $\Sigma_{\rm H2}$ show no comparable limiting values.

The vertical dashed line in the left and the right columns of Figure
\ref{fig-2} shows $\Sigma_{HI,{\rm saturation}} \approx
9$~M$_{\odot}$~pc$^{-2}$.
In Section \S\,\ref{saturation-comb} we will see that $95\,\%$ of the
$\Sigma_{HI}$ values for the combined distribution of all 7 spiral galaxies in
our sample are below $\Sigma_{HI,{\rm saturation}}$.

A second effect is best seen in the radial profile points in the left column
of Figure \ref{fig-2}: at high SFRs there is often an {\em anti-}correlation
between $\Sigma_{\rm HI}$ and $\Sigma_{\rm SFR}$. This occurs in the central
\hi -holes of spirals where the gas is overwhelmingly molecular and the star
formation rate is very high. All 7 of our spiral galaxies show some degree of
this effect, i.e., at least a mild central depression in \hi.

\subsection{$\Sigma_{\rm SFR}$ Versus $\Sigma_{\rm gas}$ in \hi-dominated
  Galaxies}
\label{dwarf_individ}

\begin{figure*}
\plotone{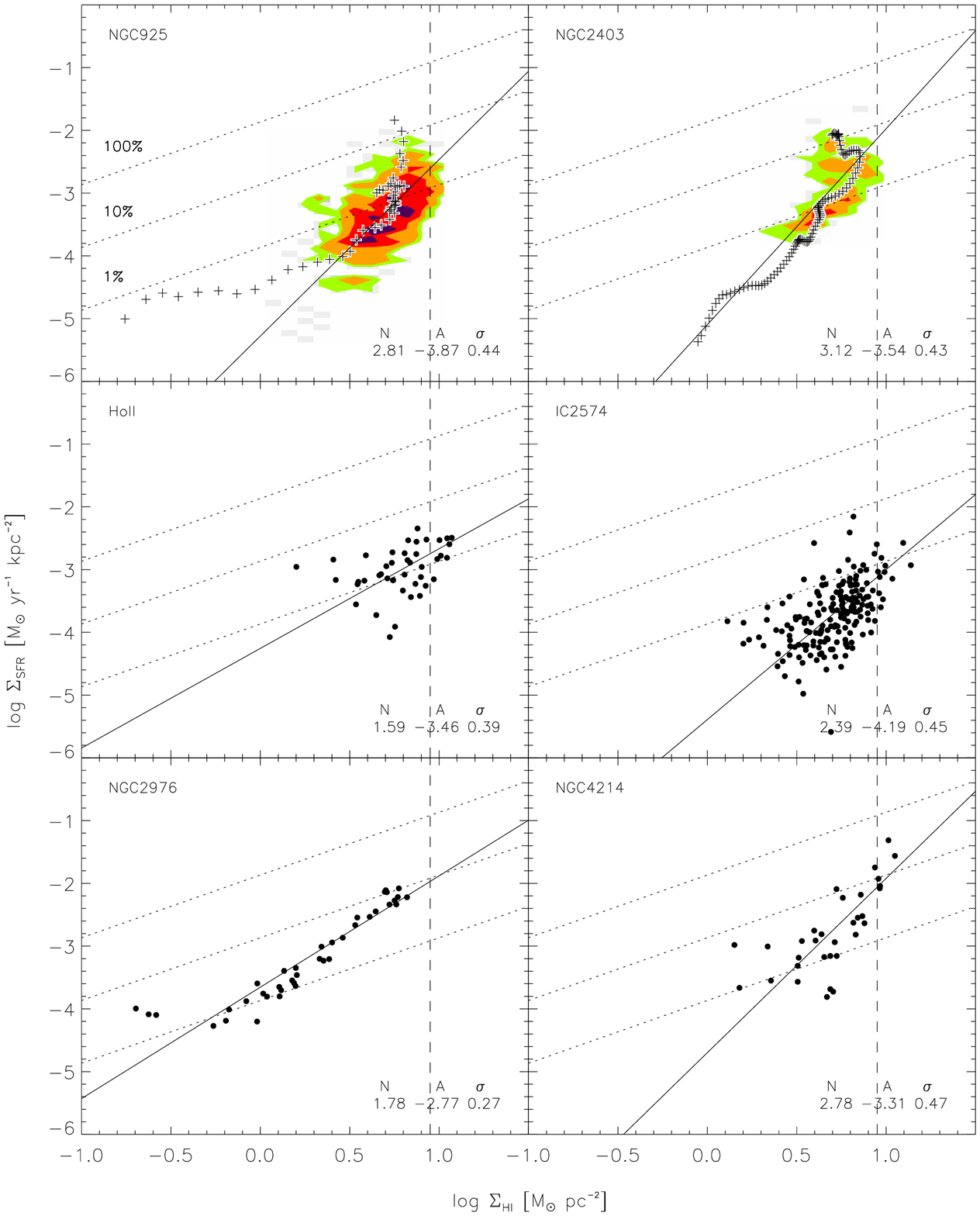}
\caption{Contour and scatter plots respectively of $\Sigma_{\rm SFR}$ versus $\Sigma_{\rm HI}$
  for 2 \hi-dominated spirals and $4$ dwarf irregular galaxies at
  750~pc resolution. Contour levels, color coding and the dotted diagonal lines of constant SFE are
  identical to Figure \ref{fig-2}. For NGC~925 and NGC~2403, crosses show average
  measurements over tilted rings from the radial profiles. For the 4 dwarf irregulars we show
  scatter plots instead of contours due to the lower number of available sampling points.
  The dashed vertical line indicates the $\Sigma_{\rm HI}$ saturation level that was
  found for the spirals.}
\label{fig-3}
\end{figure*}

We have already seen that there are variations in the relationship
between $\Sigma_{\rm gas}$ and $\Sigma_{\rm SFR}$ among spiral
galaxies, mainly in the \hi-dominated parts. Figure \ref{fig-3} shows
the relationship between $\Sigma_{\rm SFR}$ and $\Sigma_{\rm HI}
\approx \Sigma_{\rm gas}$ for 6 \hi-dominated
galaxies. Color contours for the two largest galaxies are coded as in Figure
\ref{fig-2}. For these two galaxies we plot points from the radial profiles
(Figures \ref{fig-11} and \ref{fig-12}) on the same plot as black crosses.
Because the other 4 dwarf galaxies are small, we show
scatter plots instead of density contours.  Figure \ref{fig-3} shows
individual plots for Ho~II, IC~2574, NGC~2976 and NGC~4214. Note that
the remaining galaxies, Ho~I,
DDO~154, DDO~53, M81~DwA and M81~DwB are so small that they yield only
$1$ -- $10$ sampling points each.  We include these data only later in
Figure \ref{fig-8}, which shows aggregate data for all of our dwarf
irregular galaxies.

Figure \ref{fig-3} shows that these galaxies display the same saturation
value for $\Sigma_{\rm HI}$ as the large, centrally H$_{2}$-dominated spirals. This is somewhat surprising, as one would expect
conditions in the ISM of many of these galaxies to be less favorable
to the formation of H$_2$ from \hi\ because of comparatively low
metallicities (and thus lower dust content), shallow potential wells, lower gas densities,
and more intense radiation fields. One might therefore have expected large
reservoirs of \hi\ to survive in these galaxies at columns where the
ISM is mostly molecular in a spiral.

For the most part these galaxies show low SFEs and a steep
distribution of $\Sigma_{\rm SFR}$ as a function of $\Sigma_{\rm
  HI}$. This includes IC~2574, Holmberg~II, and all of the `small'
irregulars not plotted. The notable exception is NGC~2976, which shows
an SFE similar to that found in spiral galaxies and a clear
relationship between $\Sigma_{\rm SFR}$ and $\Sigma_{\rm gas}$. As
with NGC~4736, this is driven largely by the presence of low
gas surface densities in NGC~2976 that are rarely found in the
optical disks of many of the other dwarf galaxies. NGC~2976 shows
a steadily declining \hi\ profile, perhaps curtailed by interactions
with other members of the M81 group.

We know from mapping and single dish measurements that these galaxies are not
H$_2$-dominated (for a Galactic conversion factor), so $\Sigma_{\rm HI}$ is
likely to be a good proxy for the total gas over most lines of sight.
However, if the CO-to-\htwo\ conversion factor were to vary dramatically, as has
been suggested for dwarf irregular galaxies \citep[e.g.,][]{madden97,israel97,leroy07}, or $\Sigma_{\rm H2} > \Sigma_{\rm HI}$ locally, then $\Sigma_{\rm HI}$ may
severely underestimate $\Sigma_{\rm gas}$ towards the star forming peaks. The
topic is too complex to address here, but we note the sense of the
uncertainty: if $\Sigma_{\rm gas} > \Sigma_{\rm HI}$, then the points in
Figure \ref{fig-3} will move to the right, to higher gas surface densities for
the same $\Sigma_{\rm SFR}$. Most dwarf irregulars already show lower SFEs
than most spiral galaxies; the inclusion of a substantial reservoir of
molecular gas would lower the SFE further, thus widening the difference.

\subsection{Dependence on Resolution}
\label{sflaw-scale}

\begin{figure*}
\epsscale{0.7}
\plotone{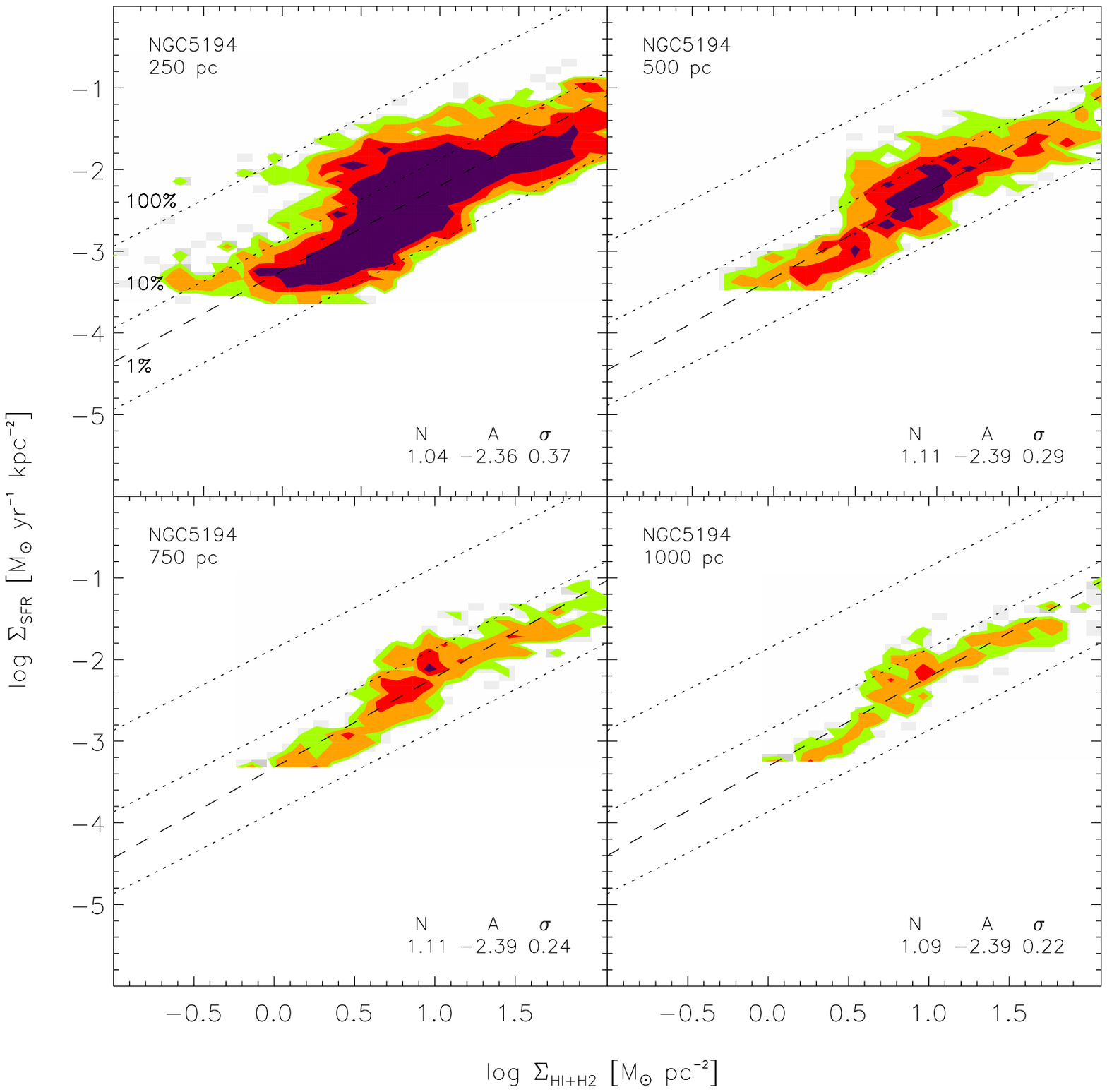}
\caption{Contour plots of $\Sigma_{\rm SFR}$ versus
  $\Sigma_{\rm gas}$ for NGC~5194 (M51) and NGC~6946 at 4 different spatial
  resolutions: the individual maximum resolution, and then 500, 700 and
  1000\,pc. Contour levels and other features of the plots, including the diagonal dotted
  lines of constant SFE, are
  identical to those in Figure \ref{fig-2}. Although the number of
  independent data points dwindle, we do not see the distribution
  change markedly with resolution.}
\label{fig-13}
\end{figure*}

\begin{figure*}
\figurenum{\ref{fig-13}}
\epsscale{0.7}
\plotone{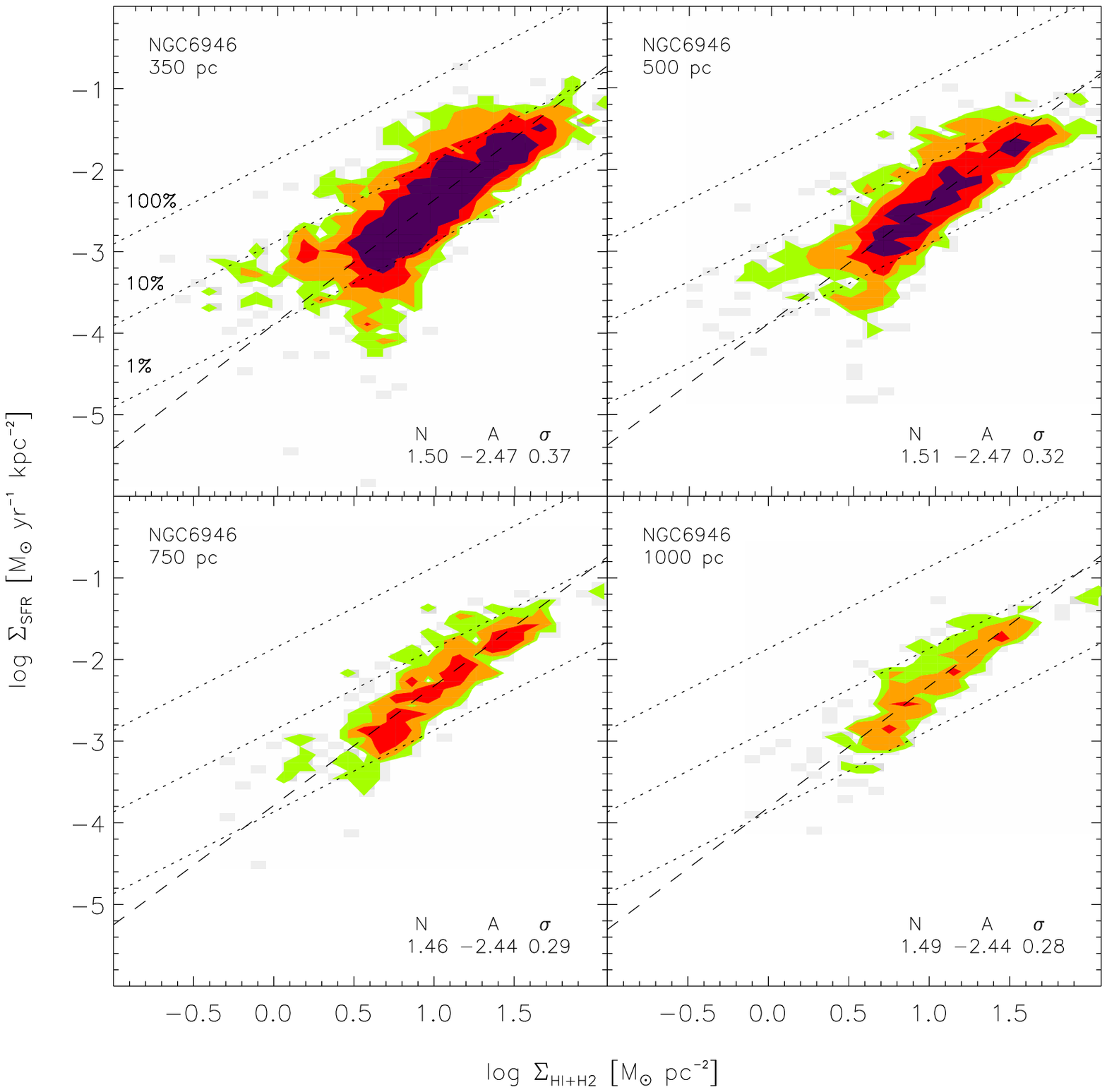}
\caption{continued.}
\end{figure*}

We have so far considered the relationship between gas and star
formation at 750\,pc spatial resolution. Cloud formation, stellar
feedback, and indeed a breakdown in our SFR tracers may all be strong
functions of spatial scale.  Therefore we investigate how the
relationships between $\Sigma_{\rm SFR}$ and $\Sigma_{\rm gas}$ or
$\Sigma_{\rm H2}$ change with spatial scale. To do this, we convolve
all of our data to a variety of spatial resolutions (see \S \ref{alignment})
spanning the range from the
original (highest) resolution to 10~kpc and repeat the above analysis.

Figure \ref{fig-13} shows the distribution of $\Sigma_{\rm gas}$
versus $\Sigma_{\rm SFR}$ for a range of spatial resolutions in two
spiral galaxies, NGC~5194 (M51) and NGC~6946. We show $\Sigma_{\rm
  SFR}$ versus $\Sigma_{\rm gas}$ for 4 spatial resolutions, the
original (best) spatial resolution, and then 500~pc, 750~pc and 1~kpc. Contours
and other details are identical to Figure \ref{fig-2}. Aside from the expected
decrease in the number of independent sampling points and some
narrowing of the distribution as a result of averaging, we do not observe any
strong effect due to spatial resolution.

\begin{figure*}
\plotone{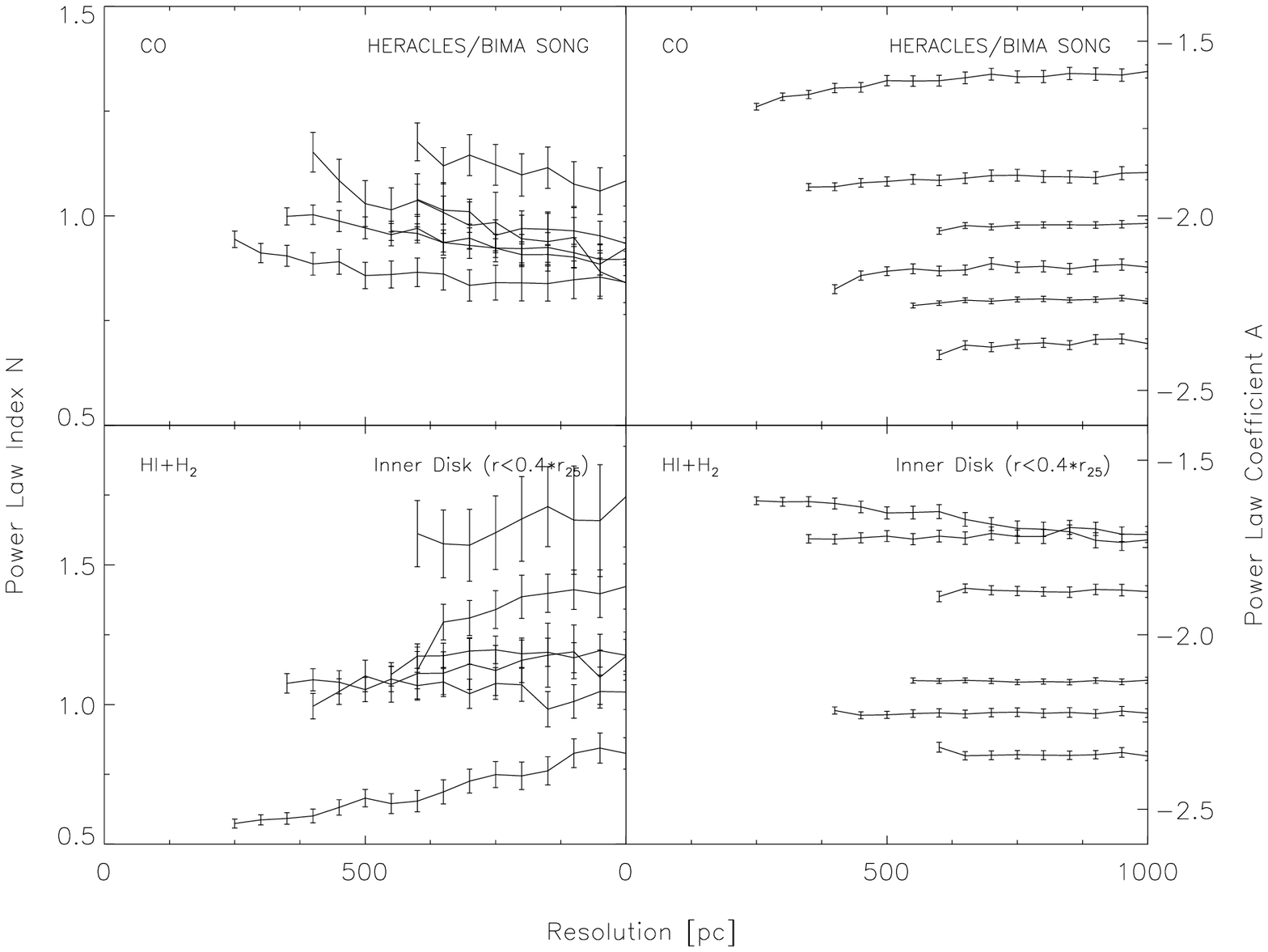}
\caption{Power law fit parameters, power law index $N$ and coefficient
  $A$ (see Equation \ref{power-law}), for 6 of our 7 spiral galaxies (NGC~4736 is omitted). Left
  panels: $N$ vs. spatial resolution. Right panels: $A$ vs spatial
  resolution. The top panels show fits to HERACLES/BIMA SONG CO data, the bottom
  panels to the total gas data within $0.4\,r_{25}$. Both power law parameters vary
  only weakly with changing spatial resolution.}
\label{fig-14}
\end{figure*}

To look more quantitatively at the effects of resolution on our
results, we fit power laws to the data at each resolution. As noted before, a single
power law is an inadequate description of the data for many galaxies;
the fits here, however, serve as a shorthand to characterize
the data distributions. Figure \ref{fig-14} shows obtained power
law parameters as a function of resolution for 6 of our 7 spiral galaxies
(star formation in NGC~4736 is too
concentrated for this exercise). The fits become more uncertain as the number of
sampling points dwindle and averaging substantially lowers our dynamic
range for many galaxies. Therefore we can carry out this test between
the best available resolution and a spatial resolution of
1~kpc. The left panels in Figure \ref{fig-14} show the power law
indices, $N$, as a function of resolution; the right panels show the
power law coefficients, $A$. For the upper panels, we fit all CO data
above the respective sensitivity limit (see \S\,\ref{co}). For the bottom
panels, we fit all total gas data within $0.4~r_{25}$.

Figure \ref{fig-14} shows that the derived fits change slowly as the
spatial resolution decreases from $200-500$~pc to $\sim 1$~kpc.
If stellar feedback and cloud formation exert a strong
influence on the relation between star formation and neutral gas (and
indeed one would expect them to) then they do so on scales $\lesssim
300$~pc. This finding agrees with results from \citet{tamburro08}, who
measured the offset between \hi\ and star forming peaks near spiral
density waves in some of the same datasets we study. They found a
typical offset of $\sim 100$~pc or less between regions of star
formation and peaks in the \hi\ maps.

\section{Combined Distributions}
\label{combined}

In the previous section we saw that the relationship between
$\Sigma_{\rm gas}$ and $\Sigma_{\rm SFR}$ varied among spiral galaxies
and between spirals and \hi-dominated galaxies. We saw that the value
at which the atomic gas saturates appears constant across our entire
sample and we found a well-defined power law index relating
$\Sigma_{\rm H2}$ and $\Sigma_{\rm SFR}$. In this section we collapse
the individual contour plots in Figure \ref{fig-2} into combined
pixel--by--pixel plots for all of the galaxies in our sample in order to be able to draw general conclusions
about the star formation law in our galaxy sample.

\begin{figure*}
\plotone{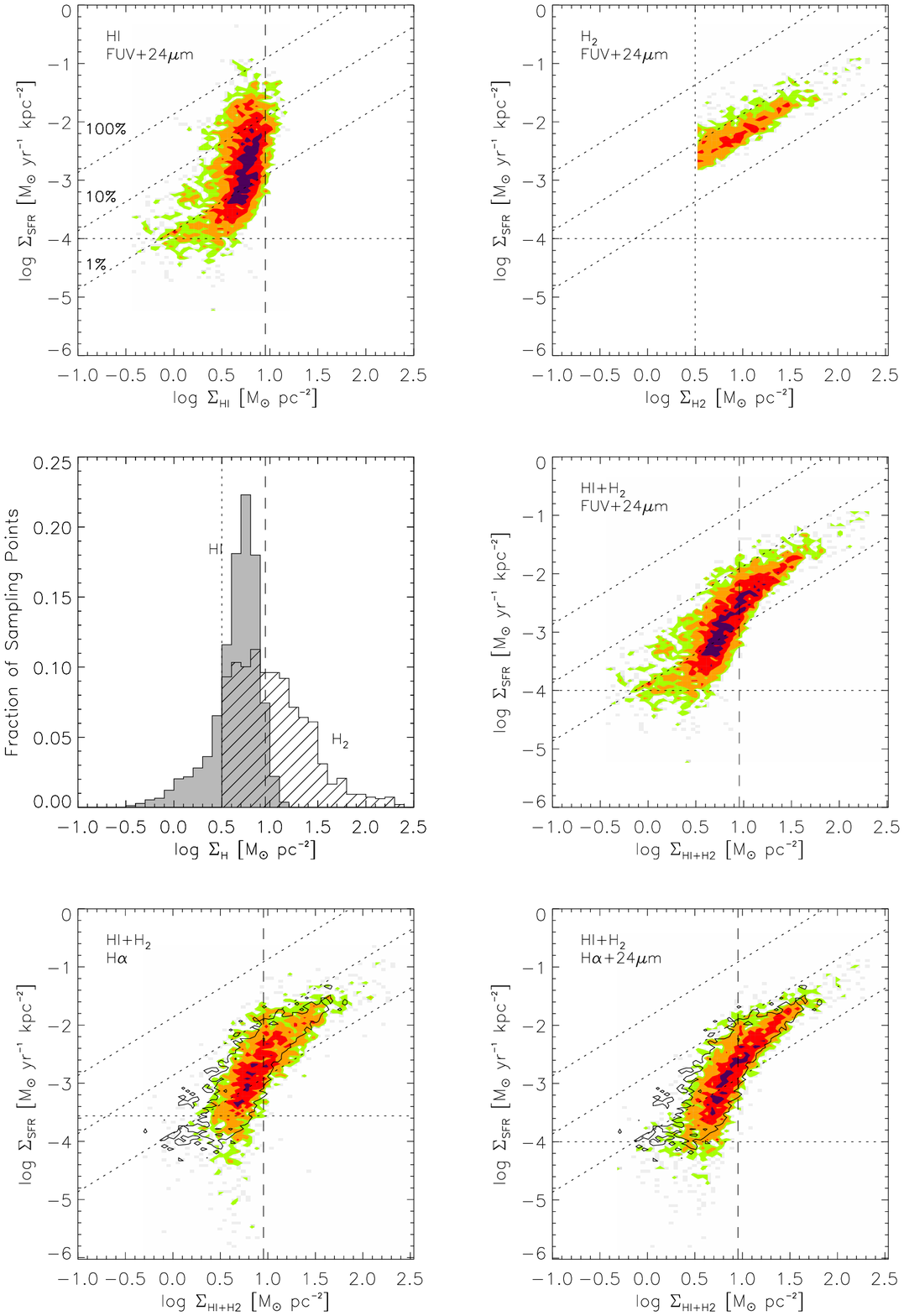}
\caption{Sampling data for all 7 spiral galaxies plotted together. Top left:
  $\Sigma_{\rm SFR}$ vs. $\Sigma_{\rm HI}$; top right: $\Sigma_{\rm SFR}$ vs.
  $\Sigma_{\rm H2}$; middle right: $\Sigma_{\rm SFR}$ vs. $\Sigma_{\rm gas}$.
  The bottom left and right panels show $\Sigma_{\rm SFR}$
  vs. $\Sigma_{\rm gas}$ using \halpha\ and a combination of \halpha\ and
  24\,\microm\ emission as SF tracers, respectively (for a subsample of 6 spirals). The sensitivity
  limit of each SF tracer is indicated by a horizontal dotted line. The black contour
  in the bottom panels corresponds to the orange contour in the middle right panel and is shown
  for comparison. The vertical dashed lines indicate the value at which $\Sigma_{\rm HI}$ saturates
  and the vertical dotted lines (top right and middle left panel) represent the sensitivity limit of the
  CO data. The diagonal dotted lines and
  all other plot parameters are the same as in Figure \ref{fig-2}. The middle left panel
  shows histograms of the distributions of \hi\ and \htwo\ surface densities (normalized to the total
  number of sampling points above the respective sensitivity limit) in the sample.}
\label{fig-4}
\end{figure*}

Figure \ref{fig-4} shows the distribution of $\Sigma_{\rm SFR}$ vs.
$\Sigma_{\rm HI}$ (top left), $\Sigma_{\rm H2}$ (top right), and
$\Sigma_{\rm gas}$ (middle right) for all sampling points in our 7
spirals. We also assess the impact of our specific choice of star formation
tracer. The bottom two panels in Figure \ref{fig-4} show the
relationship between $\Sigma_{\rm SFR}$ and $\Sigma_{\rm gas}$ using
\halpha\ (bottom left) and, alternatively, a combination of \halpha\ and 24\,\microm\
emission following \citet[][]{CALZETTI07} but here applied
pixel-by-pixel (bottom right). The bottom two plots include data for 6 spiral
galaxies: NGC~628, NGC~3184, NGC~3521, NGC~5055, NGC~5194, and
NGC~6946. All plot parameters are identical to those in Figure \ref{fig-2}.
In order to turn the \halpha\ and the combination of \halpha\ and 24\,\microm\
emission into SFRs, we assume the same IMF we use for our FUV and 24\,\microm\ based SFR maps.
The H$\alpha$-only plot furthermore includes a correction for 1.1~magnitudes of extinction for every sampling point \citep[a typical value for integrated galaxy disks,][]{kennicutt-rev98}.
The dotted horizontal line in all panels indicates the estimated sensitivity limits for $\Sigma_{\rm SFR}$. For ease of comparison, the orange contour from the middle right plot
is overplotted as the black contour on both bottom plots.

One finds that the bottom two distributions in Figure \ref{fig-4}, which are based on different
SF tracers, agree very well with the distribution that represents the SF tracer used throughout this paper (black contour in bottom panels and middle right panel). That is, our choice of star formation tracer does not appear to significantly affect the derived relationship between $\Sigma_{\rm SFR}$ and $\Sigma_{\rm gas}$.

\subsection{\hi\ Saturation in the Combined Distribution}
\label{saturation-comb}

The top left panel of Figure \ref{fig-4} shows $\Sigma_{\rm SFR}$ vs.
$\Sigma_{\rm HI}$. This plot clearly demonstrates the saturation effect
discussed for the individual galaxies (see \S\,\ref{saturation}) and shows that it is a universal feature in our sample of spirals. The middle left plot shows normalized histograms of $\log \Sigma_{\rm
  HI}$ and $\log \Sigma_{\rm H2}$. The \hi\ shows a clear truncation
near the threshold. By contrast, the \htwo\ shows no such
cutoff. This indicates that the \hi\ saturation corresponds to a phase transition from an atomic to a molecular ISM. The \hi\ saturation value we quote
of $\Sigma_{HI,{\rm saturation}} \approx 9$~M$_{\odot}$~pc$^{-2}$ represents the 95th percentile of the
\hi\ distribution. This value coincides with the \hi\ saturation value
derived from \citet{wong02} using radial profiles in a sample of 6
molecule-rich spiral galaxies and agrees with our own radial profiles.

\subsection{\hi , \htwo , Total Gas, and the Star Formation Law}

The top two panels in Figure \ref{fig-4} show that
$\Sigma_{\rm HI}$ and $\Sigma_{\rm H2}$ relate very differently to
$\Sigma_{\rm SFR}$. In the top left plot, a narrow range
of $\Sigma_{\rm HI}$ corresponds to a large range of $\Sigma_{\rm
  SFR}$ and/or SFE. Over less than one order of magnitude in $\Sigma_{\rm
  HI}$, $\Sigma_{\rm SFR}$ covers $\sim 3$ orders of magnitude and the
SFE spans $2$ orders of magnitude. $\Sigma_{\rm HI}$ therefore {\em cannot} be
used to predict either $\Sigma_{\rm SFR}$ or the SFE in spiral
galaxies. By contrast, the top right panel shows that $\Sigma_{\rm H2}$
exhibits a clear, monotonic relationship with $\Sigma_{\rm SFR}$ with a
slope of approximately unity (see \S\,\ref{sflaw-molecular}) down to the sensitivity limit of our CO data.

It is clear from Figure \ref{fig-4}
that a single power law {\em can} describe the relationship between $\Sigma_{\rm
  SFR}$ and $\Sigma_{\rm H2}$ while the distributions of $\Sigma_{\rm SFR}$ and $\Sigma_{\rm gas}$ show
a clear `knee' at the transition from an \hi\ to an
H$_2$-dominated ISM.

\subsection{The Combined Molecular Schmidt Law}
\label{sflaw-molecular}

\begin{figure*}
\plotone{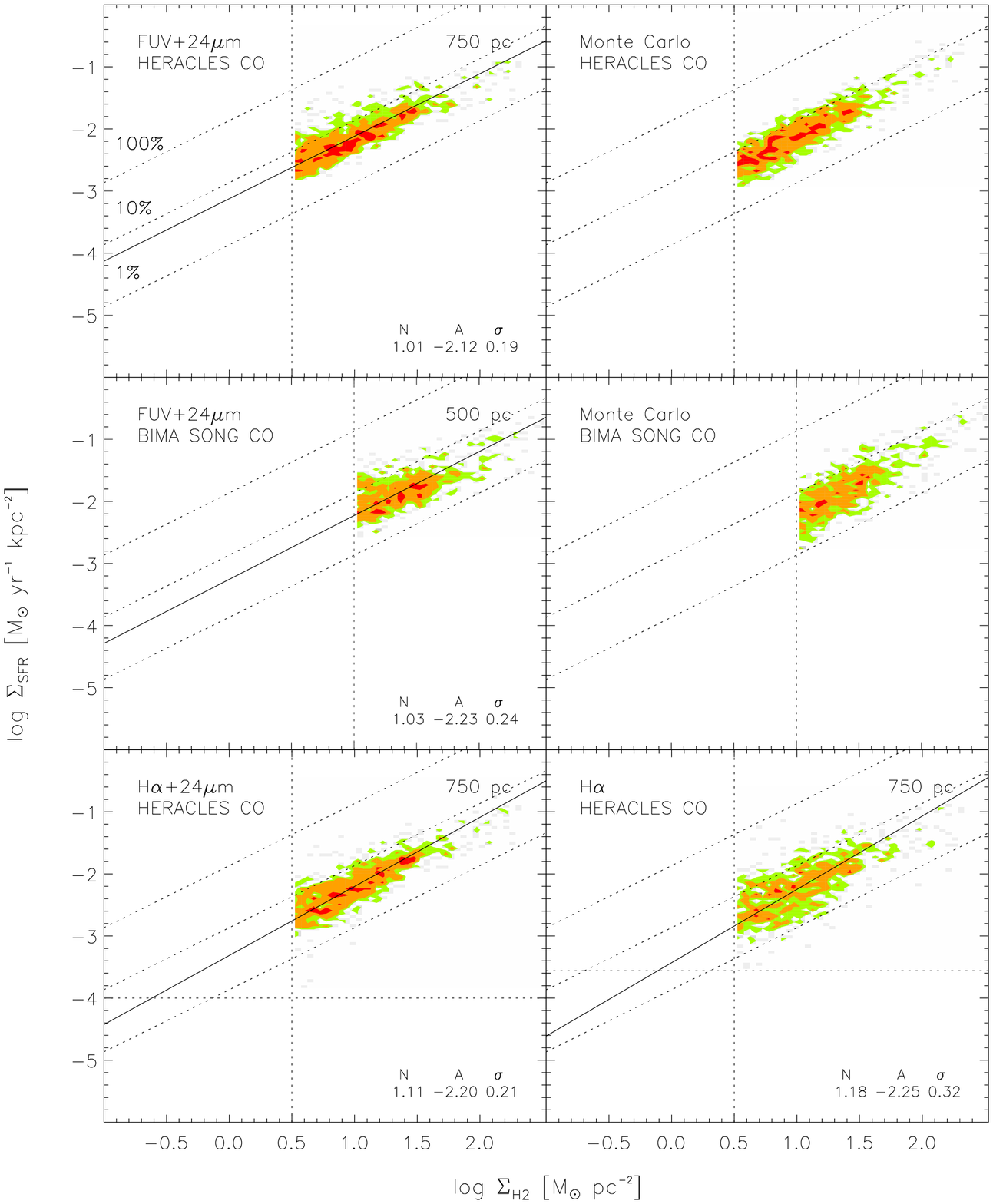}
\caption{The molecular Schmidt law ($\Sigma_{\rm SFR}$ versus $\Sigma_{\rm
    H2}$) in spiral galaxies. Green, orange and red cells
  show contours of 1, 2 and 5 data points per cell. The black solid line shows
  the best fit power law for every panel. The power law fit parameters are given
  in every panel. All other lines
  are the same as in Figures \ref{fig-2} and \ref{fig-4}. The top and middle
  right hand panels show Monte Carlo realizations of the best fit power laws using
  data values for $\Sigma_{\rm gas}$, observed scatter and best fit parameters. All other
  panels show data distributions.}
\label{fig-9}
\end{figure*}

We saw above that the common features in our data are the
saturation of the \hi\ and the linear relationship between $\Sigma_{\rm
  SFR}$ and $\Sigma_{\rm H2}$. Here we examine this `molecular Schmidt
law' for the data from all spiral galaxies combined. In the top left panel of Figure \ref{fig-9} we show the results of a power law fit to $\Sigma_{\rm SFR}$ vs. $\Sigma_{\rm H2}$ for CO data
from HERACLES/BIMA SONG in spiral galaxies (where $\Sigma_{\rm gas}
\approx \Sigma_{\rm H2}$). Contour levels and other plot parameters
are identical to the ones in Figure \ref{fig-2}.
The power law fit is shown as a solid black line. The
best fit parameters are $N = 1.01$ and $A = -2.12$; the RMS scatter of
$\Sigma_{\rm SFR}$ about the fit is $0.19$~dex, a factor of $\sim
1.5$. In this regime, our data suggest a direct proportionality
between $\Sigma_{\rm SFR}$ and $\Sigma_{\rm H2}$. The derived mean H$_{2}$ gas depletion
time (including helium) is $2.0 \cdot 10^9$~yrs with an RMS scatter of $0.8 \cdot 10^9$~yrs.

In the middle left panel the CO data comes from \bima\ alone.
This allows us to plot the data at a slightly higher resolution of 500~pc
and check the robustness of our results to changing CO maps and resolution. All other
data and plot parameters are identical to the top left panel. We
obtain an identical slope compared to that in the top left panel,
$N\,=\,1.03$ versus $N\,=\,1.01$, with a similar coefficient and scatter.

The top and middle right panels show Monte Carlo realizations, where we compute
$\Sigma_{\rm SFR}$ using $\Sigma_{\rm gas}$ from the data and the fit parameters quoted in the
left panels including the measured lognormal scatter. The agreement
between the left and right panels provides a qualitative check that a
single power law is a good description of our data in this regime.

The bottom left and right panels show how the choice of star formation tracers
affects the derived molecular Schmidt law. We use the HERACLES/BIMA SONG maps (at
750\,pc resolution) and replace our $\Sigma_{\rm SFR}$ maps with maps
derived from \halpha\ emission (bottom left plot) and a combination of \halpha\ and 24\,\microm\ emission \citep[bottom right plot,][]{CALZETTI07}. The \halpha\ includes a 1.1 magnitude correction for internal extinction. For these data we obtain $N\,=\,1.11$ and
$N\,=\,1.18$ respectively. As in Figure \ref{fig-4}, the choice
of a specific SFR tracer has only marginal impact on our results.

We thus derive a best fit molecular Schmidt law of 
\begin{equation}
\Sigma_{\rm SFR} = 10^{-2.1 \pm 0.2} \Sigma_{\rm H2}^{1.0 \pm 0.2}.
\end{equation}
\noindent for the ensemble, identical to within the uncertainties
to $N = 1.0 \pm 0.1$ and $A = -2.1 \pm 0.2$ found for individual galaxies (see \S\,\ref{fits-individual} and \S\,\ref{mol-gas-schmidt-indiv}). The uncertainties of 0.2 take into account variations in SF tracers, substitution of CO maps and scatter in the data. They do not reflect variations in the CO-to-H$_{2}$ conversion factor, the IMF or systematics in our methodology.

\section{The Star Formation Law and Environment}
\label{environment}

We saw above that there are strong variations in the SFE within many
spiral galaxies and between spirals and \hi-dominated galaxies. In
this section we attempt to link variations in the $\Sigma_{\rm
  SFR}$-$\Sigma_{\rm gas}$ relation outside the H$_2$-dominated regime in the spirals
to variations in environment. We approach this in two ways: by
measuring how the SFE changes with radius within
spiral galaxies and by comparing \hi-dominated galaxies to spirals.

\subsection{The Radial Dependence of the SFE}
\label{radial-sfe}

Our data show that below $\Sigma_{\rm gas} \approx
9$~M$_{\odot}$~pc$^{-2}$, the efficiency with which gas forms stars
varies dramatically at a given gas surface density. An obvious
explanation is that star formation thresholds of the sort discussed
by, e.g., \citet{KENNICUTT89}, \citet{martin01}, \citet{schaye04} or \citet{leroy08}
are affecting the relationship between gas and star formation. We
sample out to $r_{25}$ and \citet{martin01} find that the last \hii\
region often falls within or near this radius. The thresholds
described by various authors stem from a variety of physics. Shear,
Coriolis forces, pressure, metallicity, and passage through spiral
arms may all play key roles in the formation of gravitationally bound
molecular clouds, the necessary prerequisite to star formation.

\begin{figure*}
\plotone{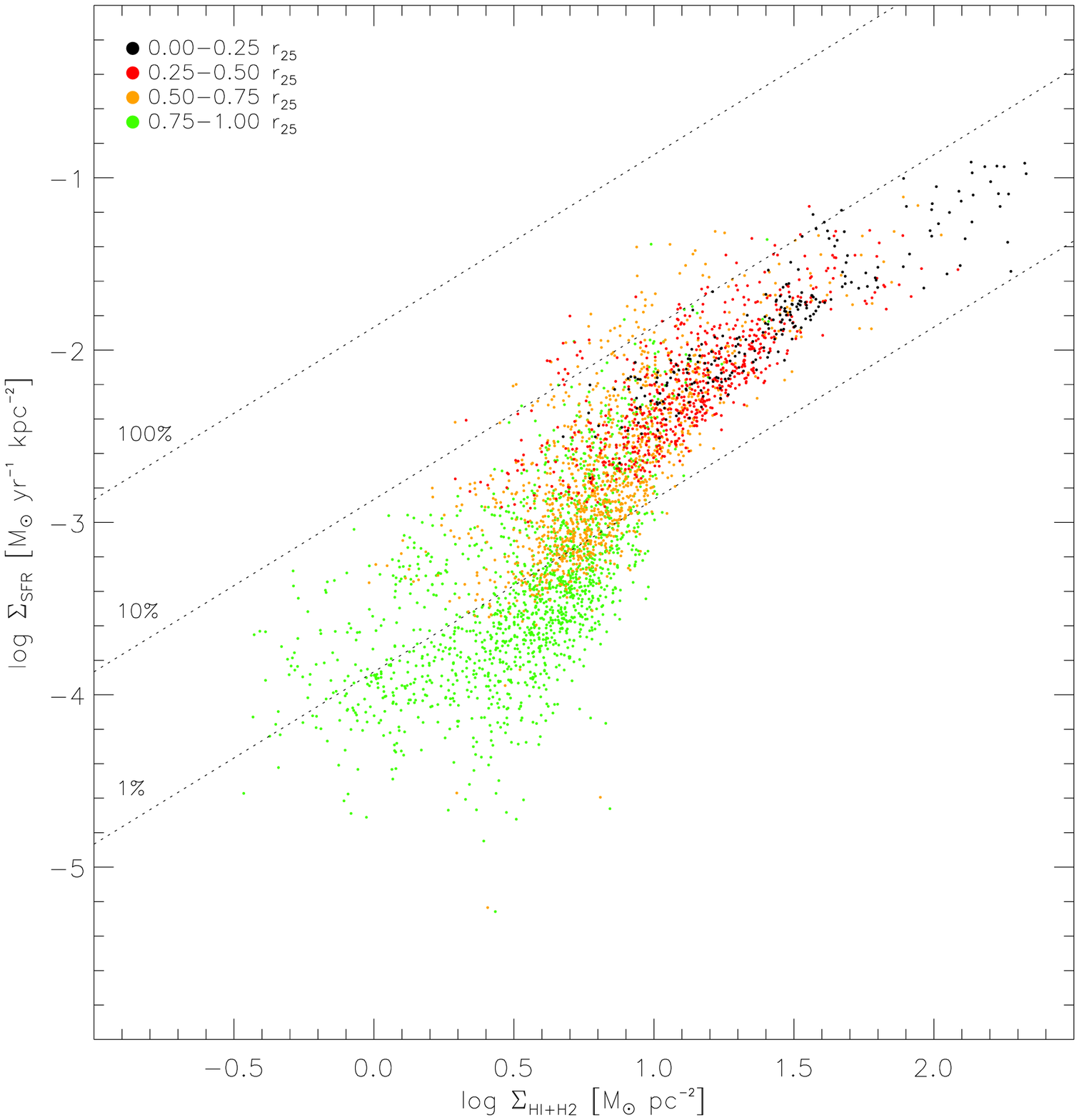}
\caption{$\Sigma_{\rm SFR}$ versus $\Sigma_{\rm gas}$ for the spiral
  galaxies with data colored according to galactocentric radius.  Every data
  point represents one sampling point; the data are the same as in the middle right plot of
  Figure \ref{fig-4}. We show the normalized
  galactocentric radius of every sampling point via its color, where, in units of $r_{25}$,
  the colors correspond to: black $< 0.25$; red
  $0.25$--$0.5$; orange $0.5$--$0.75$; and green $>0.75$. The
  diagonal dotted lines and all other plot parameters are the same as in Figure
  \ref{fig-2}. The data clearly
  break up according to galactocentric radius, with low SFE/high gas
  depletion time points corresponding to the outer parts of the spirals.}
\label{fig-6}
\end{figure*}

An exhaustive investigation of all of these other quantities is beyond the
scope of this paper but is the topic of a companion paper \citep{leroy08}.
However, most of the quantities that have been proposed
as critical to star formation vary {\em radially}. Therefore, in Figure
\ref{fig-6} we show a scatter plot $\Sigma_{\rm SFR}$ vs. $\Sigma_{\rm gas}$ for all 7 spirals.
Different colors indicate the galactocentric radius of each sampling point. The
plotted data are otherwise identical to the middle right plot in Figure
\ref{fig-4}. The deprojected galactocentric
radius of each point, normalized to the optical radius of the galaxy,
determines the color of the point: points within 0.25\,$r_{25}$ are black;
sampling points between 0.25--0.5~$r_{25}$ are red, those between 0.5--0.75~$r_{25}$ are orange; and sampling points from 0.75--1.0 $r_{25}$ are green.

Figure \ref{fig-6} shows that in the centers of spirals, where the gas is
mostly molecular, the H$_{2}$ depletion time is nearly constant (see \S\,\ref{sflaw-molecular}). At larger radii, where the ISM is dominated by \hi, the SFE of gas at a
particular surface density depends on the galactocentric radius: The further out in the disk, the lower the SFE.

\begin{figure*}
\plotone{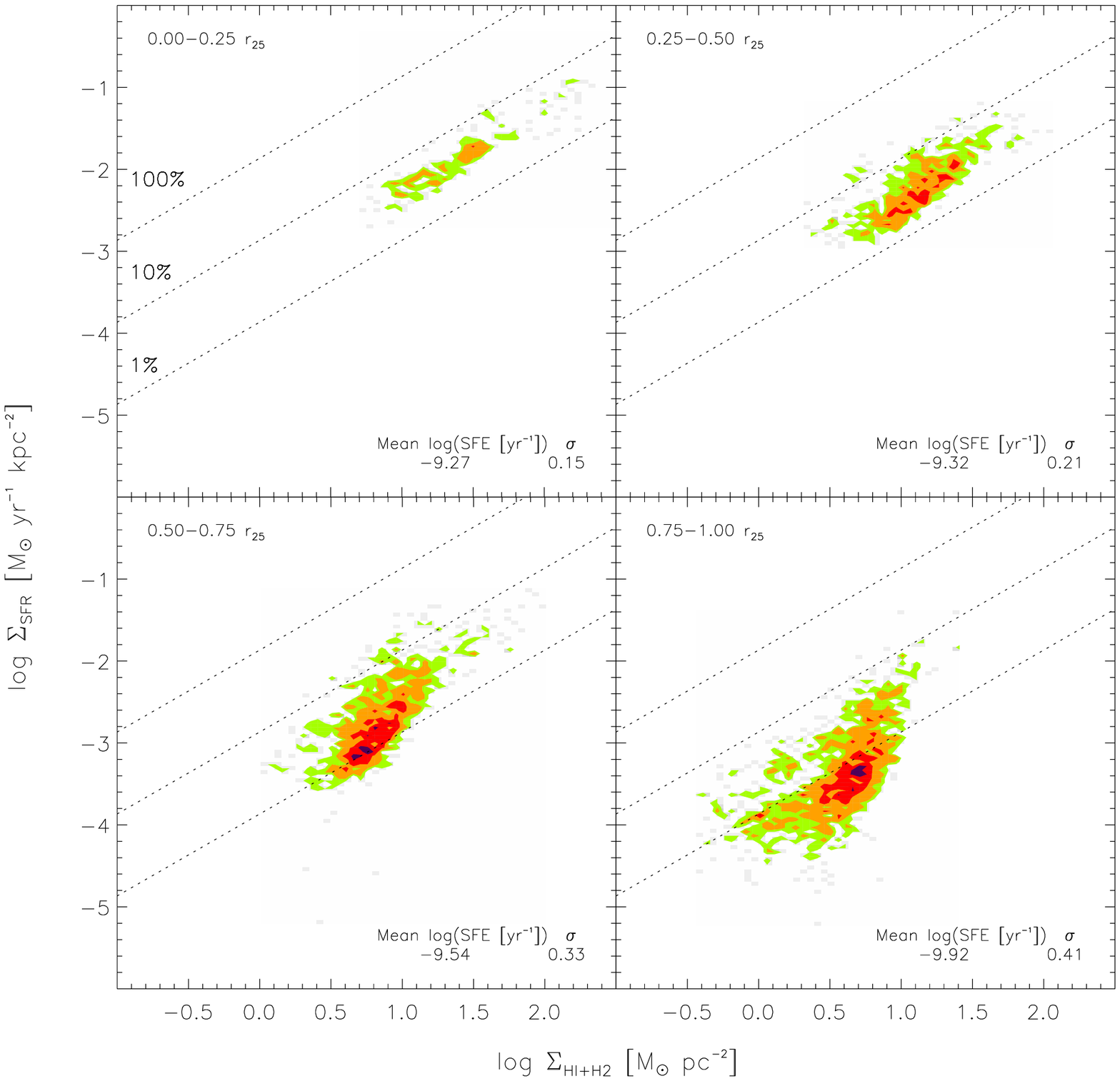}
\caption{The variation of $\Sigma_{\rm SFR}$ versus $\Sigma_{\rm gas}$
  with radius in spiral galaxies. Levels of 1, 2, 5 and 10 points per
  cell are shown as green, orange, red and magenta contours. The
  diagonal dotted lines and all other plot parameters are the same as in Figure
  \ref{fig-2}. Every plot represents only sampling data from a certain range in
  normalized galactocentric radius, corresponding to a particular
  color of points in Figure \ref{fig-6}. Upper left panel: $<0.25~r_{25}$
  (black points); upper right: $0.25$--$0.5~r_{25}$ (red points); lower
  left: $0.5$--$0.75~r_{25}$ (orange points); lower right:
  $0.75$--$1.0~r_{25}$ (green points). Data from outer galaxy disks
  display lower $\Sigma_{\rm SFR}$ for the same $\Sigma_{\rm gas}$
  compared to data from the inner disks.}
\label{fig-7}
\end{figure*}

Figure \ref{fig-7} separates the data from Figure \ref{fig-6}
into individual plots of $\Sigma_{\rm SFR}$ versus $\Sigma_{\rm gas}$
for specific ranges of galactocentric radius. We use the same radius
bins as in Figure \ref{fig-6} and show a separate contour plot for
each bin. The upper left panel shows the black
sampling points ($<0.25~r_{25}$) from Figure \ref{fig-6}, the upper
right plot shows the red points ($0.25$--$0.5~r_{25}$), the lower left plot
the orange ($0.5$--$0.75~r_{25}$), and the lower right plot shows the
green sampling points ($0.75$--$1.0~r_{25}$). In all plots, we quote
the mean SFE in that radius bin and its RMS scatter.

Figure \ref{fig-7} shows that where $\Sigma_{\rm HI} > \Sigma_{\rm H2}$,
$\Sigma_{\rm SFR}$ is a function not only of $\Sigma_{\rm gas}$, but also of
local conditions that vary strongly with galactocentric radius. These include, e.g.,
metallicity, stellar surface density, gas pressure, galactic rotation, and shear.
\citet{leroy08} explore the relationship between these quantities, $\Sigma_{\rm SFR}$,
and $\Sigma_{\rm gas}$ using the same data we use here.

\subsection{\hi-dominated Galaxies and the Outer Disks of Spirals}
\label{dwarfs}

We saw in \S\,\ref{dwarf_individ} that the SFE in \hi-dominated
galaxies is lower than that in spiral galaxies. In the
previous section, we showed that the SFE varies dramatically within a
spiral galaxy as a function of radius. Here we compare these two
findings.

\begin{figure*}
\plotone{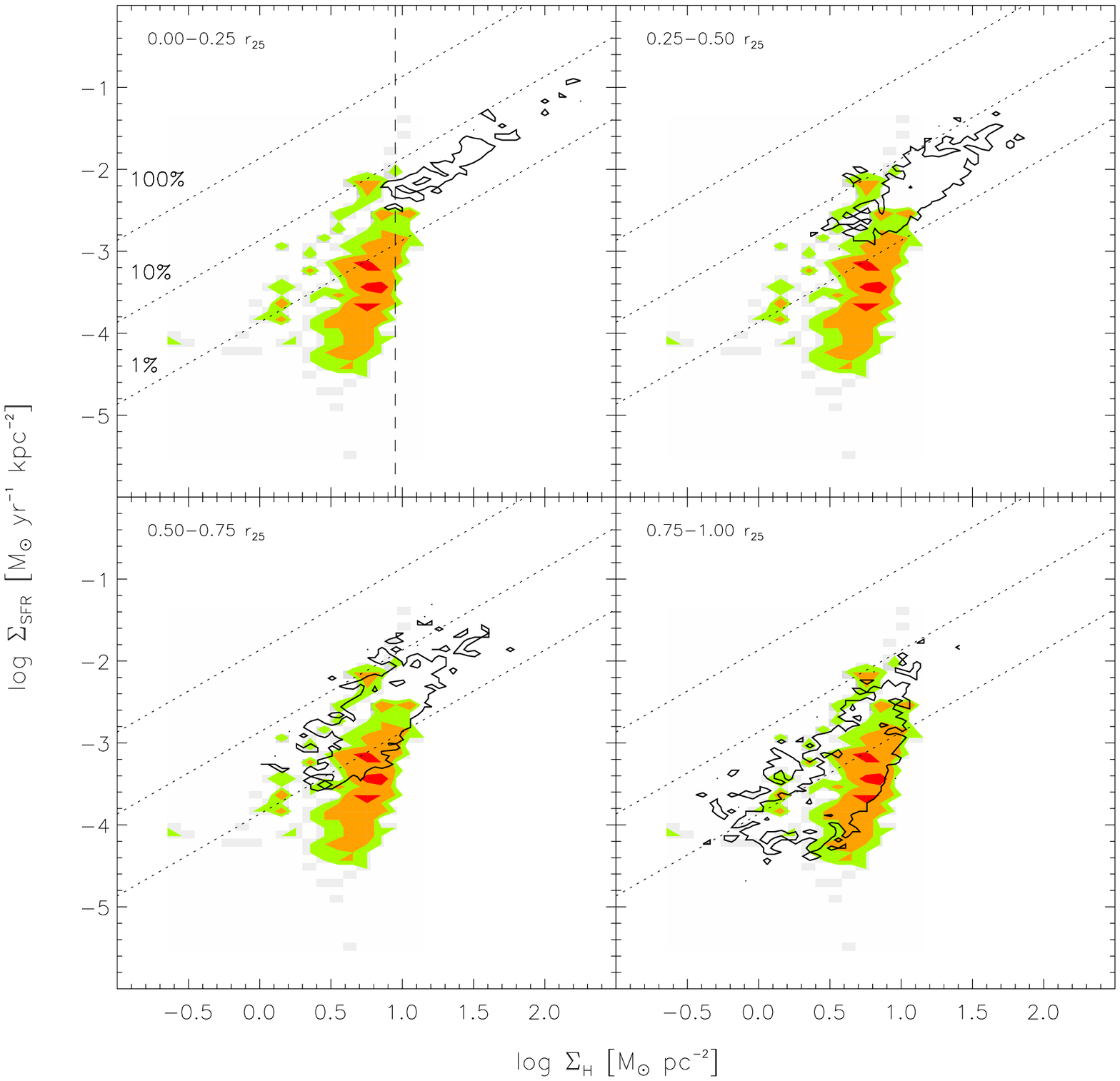}
\caption{A comparison between \hi-dominated/dwarf irregular galaxies and different
  radial regimes of spiral galaxies. All four panels show $\Sigma_{\rm SFR}$ versus
  $\Sigma_{\rm HI}$ for the dwarfs in colored contours. Green, orange, and red
  contours show 1, 2, and 5 sampling points per cell. The
  diagonal dotted lines and all other plot parameters are the same as in Figure
  \ref{fig-2}. Overplotted is the lowest (green) contour from
  the corresponding panel in Figure \ref{fig-7}. Thus each panel compares the
  distribution of data from the dwarfs to that in the spiral galaxies from a particular radial
  range. The best agreement is seen in the bottom right panel, in which the
  black contour shows data from $0.75$--$1.0$~$r_{25}$ in spiral galaxies.}
\label{fig-8}
\end{figure*}

Figure \ref{fig-8} shows the combined distribution of $\Sigma_{\rm
  SFR}$ versus $\Sigma_{\rm HI}$ for all dwarf irregular galaxies in
our sample (i.e., the \hi-dominated galaxies from Table \ref{table-general},
omitting NGC~925 and NGC~2403 that would otherwise dominate the distribution).
We show the same data for the \hi-dominated galaxies in all 4
panels. In each panel, we overplot a black contour that shows
$\Sigma_{\rm SFR}$ versus $\Sigma_{\rm gas}$ for the spiral galaxies from a
particular radial bin. This black contour in each panel of Figure \ref{fig-8}
corresponds to the lowest (green) contour of the respective panel in Figure \ref{fig-7}.

One finds that the distribution of $\Sigma_{\rm SFR}$ versus $\Sigma_{\rm gas}$ for
the \hi-dominated/dwarf galaxies overlaps the distribution of the
outer disks of the spiral galaxies (compare bottom right panel of
Figure \ref{fig-8}). Many conditions are similar in these two regimes --
low metallicities, low dust to gas ratios, high atomic gas fractions,
and comparatively weak stellar potential wells. This plot suggests
that these shared environmental factors lead to a similar relationship
between gas and star formation in both regimes.

We note that one environmental fact that dwarf irregulars do not share with
the outer disks of spiral galaxies is differential rotation. Dwarf galaxies
tend to have nearly solid body rotation curves and correspondingly low shear,
whereas rotation curves tend to be flat in the outer parts of spirals. This
implies that shear alone may not be the driving force regulating the SFE.

\subsection{The Molecular-to-Atomic Gas Ratio $\Sigma_{\rm H2}\ / \Sigma_{\rm
    HI}$ as a Function of Radius}
\label{gas-fraction}

\begin{figure*}
\plotone{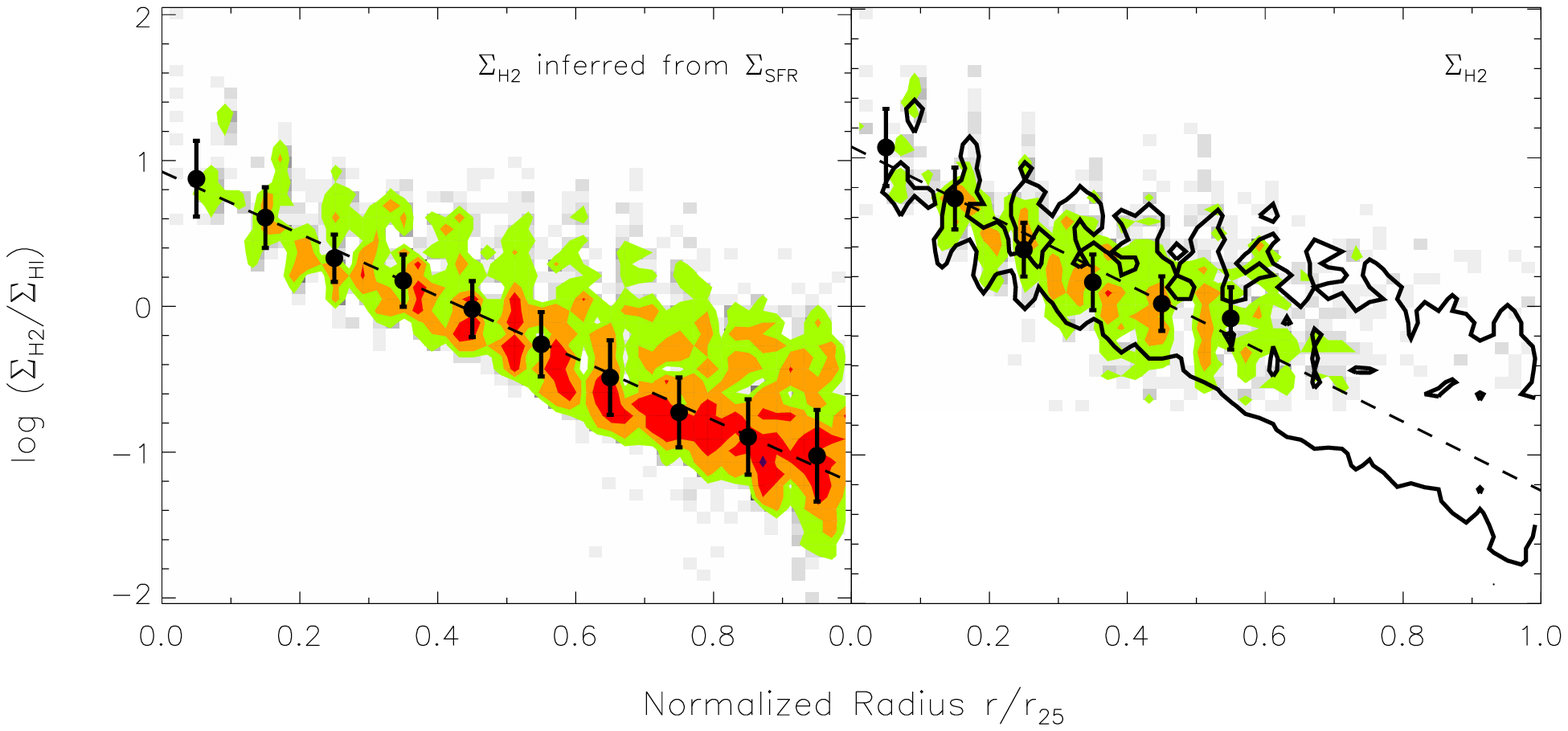}
\caption{The ratio of molecular to atomic gas, $\Sigma_{\rm H2}$ /
  $\Sigma_{\rm HI}$, as a function of galactocentric radius for the spiral
  galaxies. In the left panel, $\Sigma_{\rm H2}$ is inferred from
  $\Sigma_{\rm SFR}$ using the direct proportionality between the two
  that we have established in \S\,\ref{sflaw-molecular}. In the right panel,
  $\Sigma_{\rm H2}$ from HERACLES is used directly. The black contour in the right panel
  is identical to the green contour in the left panel and is shown for comparison.
  Green, orange, and red cells show contours of 2, 5 and 10
  sampling points per cell (cell sizes: $\Delta x = 0.02~r_{25}$, $\Delta y = 0.08$~dex). The filled
  black circles show the median
  $\Sigma_{\rm SFR}$ per $0.1~r_{25}$ radial bin, the error bars represent the
  $1 \sigma$ scatter in each bin. We carry out exponential fits to both distributions
  (dashed black lines).}
\label{fig-10}
\end{figure*}

In \S\,\ref{radial-sfe} we saw that where $\Sigma_{\rm HI}\ > \Sigma_{\rm H2}$,
the SFE varies strongly with radius. In combination with our finding that $\Sigma_{\rm SFR} \propto \Sigma_{\rm H2}$ (see \S\,\ref{sflaw-molecular}), this leads us to expect that the H$_{2}$-to-\hi\ ratio also varies strongly with radius.
The right panel in Figure \ref{fig-10} shows $\Sigma_{\rm H2} /
\Sigma_{\rm HI}$ as a function of normalized galactocentric radius for
all 7 spiral galaxies.

We find that the H$_{2}$-to-\hi\ ratio decreases as a function of radius. But the sensitivity of our CO data limits the measurement beyond $\sim$0.5\,$r_{25}$. An alternate approach is therefore to extrapolate $\Sigma_{\rm H2}$ from $\Sigma_{\rm SFR}$ using the direct proportionality between the two that we have established in \S\,\ref{sflaw-molecular} for the inner part of the
optical disk. The FUV and 24\,\microm\ data are more sensitive than
our CO maps and so offer more significant measurements in the outer disk. For this
approach, we assume that the relationship that we fit between
$\Sigma_{\rm SFR}$ and $\Sigma_{\rm H2}$ holds across the disks of our spiral
galaxies and infer the ratio $\Sigma_{\rm H2}\left( \Sigma_{\rm SFR} \right) /
\Sigma_{\rm HI}$ for each sampling point. This is shown in the left panel. For comparison, the green
contour in the left panel is overplotted in the right panel as a black contour.

One finds that the two approaches agree reasonably well; the ratio measured from CO
overlaps the ratios inferred from $\Sigma_{\rm SFR}$. Both panels show that the phase of the ISM is a well-defined function of local conditions that vary with radius. A detailed investigation is beyond the scope of this paper but we explore this topic (often referred to as `star formation thresholds') further in \citet{leroy08}. Here we show that an exponential fit, which is shown by the dashed black line in both panels, can provide a good description of the relationship between H$_{2}$, \hi\ and the SFR where $\Sigma_{\rm HI}\ > \Sigma_{\rm H2}$. The exponential fit yields a scale length of 0.2\,$r_{25}$ in both panels. The radius where $\Sigma_{\rm H2} = \Sigma_{\rm HI}$ can be derived from the coefficients of the fits. The result is identical in both cases: 0.43\,$r_{25}$ in the left panel and 0.46\,$r_{25}$ in the right panel respectively.

\section{Summary \& Discussion}
\label{summary}
\subsection{Summary of Our Work}

Using data from THINGS, HERACLES, BIMA SONG, the GALEX NGS and SINGS, we
derive $\Sigma_{\rm HI}$, $\Sigma_{\rm H2}$, and $\Sigma_{\rm SFR}$ at
750~pc resolution across the optical disks of 7 nearby spiral galaxies
with H$_2$-dominated centers and 11 \hi-dominated late-type
galaxies. We use these datasets to make the first pixel-by-pixel analysis of the
star formation law in a significant sample of nearby galaxies.

We find two relationships common throughout our data. First, a
molecular Schmidt law with index $N = 1.0 \pm 0.2$ relates $\Sigma_{\rm H2}$ to
$\Sigma_{\rm SFR}$ in our sample of spirals. This may also be described as a total
gas Schmidt law inside $\sim 0.5\,r_{25}$, where the ISM
of all of our spiral galaxies is H$_2$-dominated. The average molecular gas
(including helium) depletion time is $2.0 \cdot 10^9$ years with an RMS scatter of $0.8 \cdot 10^9$ years. This relationship holds for individual galaxies as well as for the combined
distribution and is robust to substituting BIMA SONG for HERACLES data, using
different SFR tracers, or changing the resolution from $\sim 300$~pc to $\sim 1$~kpc.

The second common feature of our data is that $\Sigma_{\rm HI}$
saturates at a surface density of $\sim 9$~M$_{\odot}$~pc$^{-2}$;
gas in excess of this value is found in the molecular
phase in the spirals. This saturation is common to spiral and \hi-dominated
galaxies. This is somewhat surprising because conditions in \hi-dominated
galaxies (such as dwarf galaxies) should be less
favorable to the formation of H$_2$, which may lead one to expect a
higher saturation value; a situation that is not observed.

We do not observe a universal relationship between total gas
surface density and $\Sigma_{\rm SFR}$. Outside the H$_2$-dominated
region, i.e., at $r \gtrsim 0.5\,r_{25}$, the relationship between gas and star formation varies both
within and among galaxies. In some cases, a single power law relates
total gas and SFR over many orders of magnitudes in gas surface
density. In other cases,
we find a wide range of star formation rates at almost the
same gas column. As a result of this variation, our best-fit power law
index, $N$, for the total gas in spiral
galaxies ranges from 1.1 to 2.7. This agrees well with the range of
indices found in the literature, but does not hint at a universal total gas
Schmidt law.

We describe variations in the relation between the SFR ($\Sigma_{\rm SFR}$) and the total gas ($\Sigma_{\rm gas}$) in terms of the star formation efficiency (SFE), i.e., star formation rate per unit gas mass ($\Sigma_{\rm SFR} / \Sigma_{\rm gas}$), and the H$_2$-to-\hi\ ratio ($\Sigma_{\rm H2} / \Sigma_{\rm HI}$). We show that outside of the central, H$_2$-dominated regime of spiral galaxies, the SFE has a strong gradient with radius, where the highest efficiency points come from the inner disk, and the lowest efficiency points are at larger radii. We also show that the pixel-by-pixel distributions of late-type, \hi-dominated
galaxies overlap those of the outer disks of the spirals in our sample in $\Sigma_{\rm SFR}$-$\Sigma_{\rm gas}$ parameter space. This implies that similar conditions, i.e.,
low metallicities, weak potential wells and low dust content, might drive the SFE in this regime.

We argue that these observations show a clear link between environment
and the relationship between gas and star formation. We suggest the
following scenario: the observational star formation `law', as it is
now understood (e.g., K98), is a molecular phenomenon. The transition
from \hi\ to \htwo\ and the subsequent formation of stars is not purely a
function of the total gas surface density. Instead, other physics sets the
ratio of \hi\ to \htwo . The critical quantity for these processes is
a strong function of radius and appears to be common to both dwarf irregular
galaxies and the outer regions of spiral disks. This agrees quantitatively with the
findings of \citet{wong02} and qualitatively with the ideas of star
formation thresholds discussed by, e.g., \citet{KENNICUTT89}, \citet{martin01},
\citet{leroy08} and many others.

We test these results in several ways and find that our
conclusions are robust against variations in the star formation rate tracer used
or the applied spatial resolution. Moreover, we find the same results
when performing a pixel-by-pixel analysis and when working with azimuthally averaged
radial profiles.

\subsection{Comparison with Previous Measurements}
\label{discussion-previous}

\begin{figure*}
\plotone{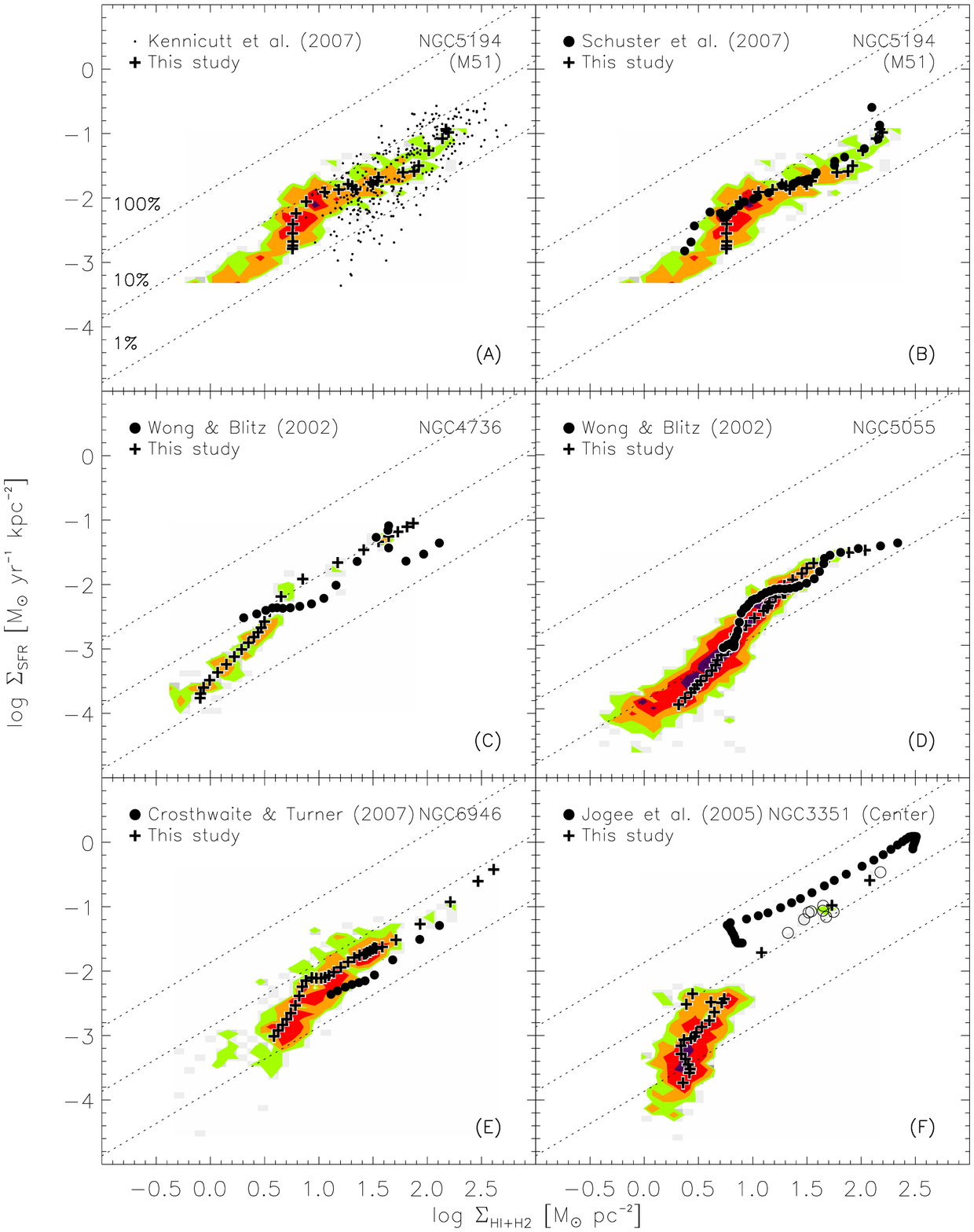}
\caption{$\Sigma_{\rm SFR}$ versus $\Sigma_{\rm gas}$ for individual galaxies from this study and from previous analyses. The colored contours represent the pixel-by-pixel data for spiral galaxies from this study; the distributions are identical to the respective ones in Figure \ref{fig-2}. The
diagonal dotted lines and all other plot parameters are the same as in Figure
\ref{fig-2}. Overplotted are datapoints from our radial profiles (compare Figures \ref{fig-11} and \ref{fig-2}) as black crosses. We compare our data to measurements from other analyses, which were carried out either by using individual apertures (panel A; each black dot represents the measurement within an individual aperture) or radial profiles (other panels; each filled black circle represents a radial profile data point). The references are given in each panel. We adjust these data to match our adopted inclinations, IMF, CO line ratio and CO-to-\htwo\ conversion factor. In general, there is good agreement between our data and the measurements from other studies, but the comparison also shows that the distributions are quite sensitive to some of the underlying assumptions, e.g., how one corrects for extinction or which CO-to-\htwo\ conversion factor is applied.}
\label{fig-15}
\end{figure*}

As we discussed in \S\,\ref{intro}, many different studies have examined the SF law in nearby galaxies and derived a variety of power law indices. These studies have used a wide range of SFR tracers, methodologies (disk averaged measurements, radial profiles or individual apertures) and assumptions (e.g., adopted IMF or CO-to-H$_{2}$ conversion factor). Here we compile the results from several studies whose galaxy samples overlap our own to compare the data distributions in $\Sigma_{\rm SFR}$-$\Sigma_{\rm gas}$ phase space once different methodologies are accounted for.

In Figure \ref{fig-15}, we therefore compare our $\Sigma_{\rm SFR}$ vs. $\Sigma_{\rm gas}$ pixel-by-pixel distributions and our radial profile data to measurements from other authors after matching the assumptions which have gone into each of their analyses to ours. In each panel, the colored contours represent the pixel-by-pixel data from our study; these distributions are identical to the ones in Figure \ref{fig-2}. The black crosses represent the datapoints from our radial profiles (Figure {\ref{fig-11}) and are also identical to the crosses in Figure \ref{fig-2}.

The black dots in Panel A and the black circles in the other panels show data from previous papers. These data were adjusted to match our assumptions regarding the adopted IMF (see \S\,\ref{sfrmaps}), the inclination, the CO-to-H$_{2}$ conversion factor, and the I$_{CO}(J=2\rightarrow1)$/I$_{CO}(J=1\rightarrow0)$ line ratio. We mention the adjustments that were made for each dataset in the discussion of the individual panels of Figure \ref{fig-15} below.

{\em Panel A}: \citet[][K07]{KENNICUTT07} derive a power law index of 1.37$\pm$0.03 relating molecular gas and SFR surface densities in M51 (and an index 1.56$\pm$0.04 for the total gas). They measure $\Sigma_{\rm gas}$ and $\Sigma_{\rm SFR}$ by placing 520\,pc apertures on \halpha\ and 24\,\microm\ emission peaks (black dots in Panel A in Figure \ref{fig-15}). K07 use Pa$\alpha$ and a combination of 24\,\microm\ and \halpha\ emission to estimate $\Sigma_{\rm SFR}$. They use the same \hi\ and CO data we do.

Shown (Figure \ref{fig-15} A) is a comparison for NGC\,5194 (M51) between our data and the measurements from K07, which we have adjusted to match our CO-to-H$_{2}$ conversion factor and IMF. One finds that our data distribution generally agrees quite well with that from K07. Nevertheless, the regime that both studies probe is slightly different: whereas we sample the entire optical disk, K07 focus on apertures mainly in the spiral arms of M51. As a consequence, our data distribution extends to lower $\Sigma_{\rm gas}$ while their distribution emphasizes slightly higher $\Sigma_{\rm SFR}$ and $\Sigma_{\rm gas}$.

The power law slope of 1.37$\pm$0.03 derived in K07 for the molecular gas is steeper than the value of 0.84 that we derive for M51 (see Figure \ref{fig-2} and Table \ref{table-fits}). The steeper slope in K07 is driven mainly by the fact that they measure low $\Sigma_{\rm SFR}$ for some apertures at log$\left(\Sigma_{\rm gas}\right)$ between $\sim$1 and 1.8 (compare black points in panel A). This difference arises because K07 subtract a local \halpha\ and 24\,\microm\ background for each aperture. This has marginal impact at high $\Sigma_{\rm SFR}$, but significantly affects low $\Sigma_{\rm SFR}$ measurements where the contrast with the local background is low. This comparison shows that despite using the same gas data and similar SFR tracers, the applied methodology for deriving SFRs and the specific choice of sampling method have non-negligible impact on the derived power law slopes.

In {\em Panel B} we compare our M51 distribution to data adopted from \citet{schuster07}, which are adjusted to match our I$_{CO}(J=2\rightarrow1)$/I$_{CO}(J=1\rightarrow0)$ line ratio, IMF and CO-to-H$_{2}$ conversion factor (we note that they apply a conversion factor equal to one quarter of our adopted value). \citet{schuster07} use radial profiles and obtain a power law index of 1.4$\pm$0.6 when fitting SFR and total gas surface densities in M51. They use radio continuum emission at 20\,cm to derive $\Sigma_{\rm SFR}$.

In this panel, we plot their data after adjusting for differences in the applied line ratio, IMF and CO-to-H$_{2}$ conversion factor and find good agreement for the distributions. The discrepancy at small $\Sigma_{\rm gas}$ arises because the \hi\ data used in \citet{schuster07} are less sensitive than our map. This leads to decreasing \hi\ columns at larger radii in their radial profile while our profile remains flat. The central radial profile point from \citet{schuster07} shows a particularly high $\Sigma_{\rm SFR}$. This may be due to the AGN in M51, which is a strong source of radio continuum emission.

{\em Panel C and D}: Both panels show radial profile data adopted from \citet{wong02}, panel C for NGC\,4736 and panel D for NGC\,5055. In both panels, the data from \citet{wong02} are adjusted to match our IMF and adopted inclinations. They derive their SFRs from \halpha\ emission and for their plotted datapoints in Panels C and D, we have adopted their constant global extinction correction of 1.5\,mag. We note that the radial profiles from \citet{wong02} only extend out to $\sim$150\arcsec\ and $\sim$300\arcsec\ for NGC\,4736 and NGC\,5055, respectively, while ours reach the edge of the optical disks ($\sim$230\arcsec\ and $\sim$350\arcsec, respectively).

For NGC\,4736, the observed $\Sigma_{\rm SFR}$ for the \citet{wong02} profile are lower than those measured by us (apart from their `bump' at log$\left(\Sigma_{\rm gas}\right) \approx$\,1.5). For this galaxy, their derived $\Sigma_{\rm SFR}$ depends sensitively on the method used to correct the \halpha\ emission for extinction: mid-IR (24\,\microm), fixed extinction, and a gas column based extinction estimate yield results that differ by a factor of about two. In the case of NGC\,5055 (Panel D), the agreement between both distributions is excellent for all $\Sigma_{\rm gas}$.

{\em Panel E} compares data for NGC\,6946 adopted from \citet{crosthwaite07}, adjusted to match our adopted inclination and IMF, to our data. \citet{crosthwaite07} use radial profiles and find a proportionality between the molecular gas and the SFR surface density (implying a power law slope of $\sim$1 for the molecular gas). They derive their SFRs from FIR and radio continuum emission. We will, however, only use the FIR based SFRs in our comparison, because it is easier to match assumptions for this case. Our radial profile extends to larger radii (and thus includes lower $\Sigma_{\rm gas}$ values) and because \citet{crosthwaite07} average emission in 30\arcsec\ wide rings (our study: 10\arcsec), they have fewer datapoints and their $\Sigma_{\rm SFR}$ and $\Sigma_{\rm gas}$ show less dynamical range. Despite an offset in $\Sigma_{\rm SFR}$, likely due to the different SFR tracers, both distributions are in good qualitative agreement.

{\em Panel F}: We compare pixel-by-pixel data from NGC\,3351, which is not part of our current sample, to inner disk data from \citet{jogee05}. Their data are adjusted to match our IMF, CO-to-H$_{2}$ factor and adopted inclination (41$\degr$). They use radial profiles to sample the inner 1\,kpc of NGC\,3351. Because they restrict their analysis to the inner disk, they only consider molecular gas. They derive their SFRs from Br$\gamma$ emission. This comparison allows us to study a regime of relatively high $\Sigma_{\rm SFR}$ and $\Sigma_{\rm gas}$. Due to the coarser resolution of our data (750\,pc) compared to the data from \citet[][120\,pc for NGC\,3351]{jogee05}, we can only compare 3 radial profile points (black crosses) and a few individual pixel-by-pixel sampling points (highlighted as black circles) to the measurements of \citet[][black filled circles]{jogee05}. We find an offset in terms of $\Sigma_{\rm SFR}$, probably due to the different SFR tracers used. Both distributions show good qualitative agreement nevertheless.

These comparisons show that once the different assumptions of all measurements are accounted for, the data distributions from the different studies agree well in general. This appears to be relatively independent of the specific SFR tracer or the sampling method (e.g., pixel-by-pixel or radial profiles) that is applied. The resolution of the data seems to matter only marginally, as, even when disk averaged measurements are included, the data still populate the same regime in $\Sigma_{\rm SFR}$-$\Sigma_{\rm gas}$ phase space (as we will discuss in \S\,\ref{discussion-regimes}).

When one corrects for different assumptions, the CO-to-H$_{2}$ conversion factor is especially important and matters particularly in the inner part of galaxies, where the molecular gas dominates the total gas budget. The factors originally applied in the above studies differ by as much as a factor of 6. If one does not correct for different conversion factors, the high $\Sigma_{\rm gas}$ end of the distributions may change quite drastically, which can significantly influence derived power law slopes.

Nuclear activity, i.e., the presence of AGN, may also play a role for radial profile datapoints from the centers of galaxies, depending on the applied SFR tracers (see discussion for Panel B). This is only a minor issue when sampling pixel-by-pixel, where only very few pixels from the center contribute.

\subsection{The Molecular Schmidt Law In Various Regimes}
\label{discussion-regimes}

\begin{figure*}
\plotone{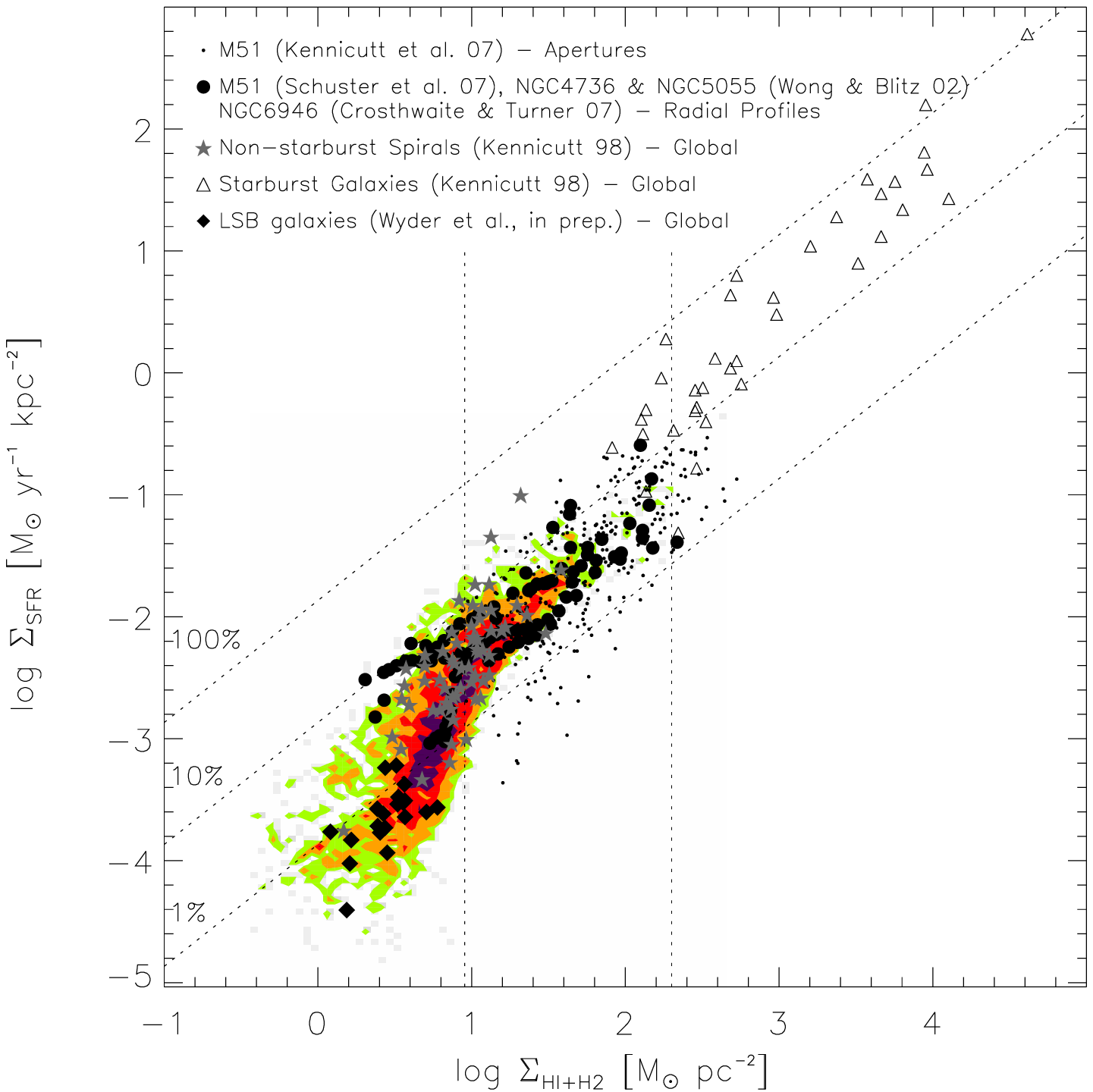}
\caption{$\Sigma_{\rm SFR}$ versus $\Sigma_{\rm gas}$ from this paper in colored contours (compare middle right panel of Figure \ref{fig-4}) and for individual galaxies from other analyses (see Figure \ref{fig-15}). The diagonal dotted lines and all other plot parameters are the same as in Figure
\ref{fig-2}. Overplotted as black dots are data from measurements in individual apertures in M51 \citep{KENNICUTT07}. Datapoints from radial profiles from M51 \citep{schuster07}, NGC\,4736 and NGC\,5055 \citep{wong02} and from NGC\,6946 \citep{crosthwaite07} are shown as black filled circles. Furthermore, we show disk averaged measurements from 61 normal spiral galaxies (filled grey stars) and 36 starburst galaxies (triangles) from K98. The black filled diamonds show global measurements from 20 low surface brightness (LSB) galaxies (Wyder et al., in prep.). Data from other authors were adjusted to match our assumptions on the underlying IMF, CO line ratio, CO-to-\htwo\ conversion factor and galaxy inclinations where applicable. One finds good qualitative agreement between our data and the measurements from other studies despite a variety of applied star formation rate tracers. This combined data distribution is indicative of 3 distinctly different regimes (indicated by the vertical lines) for the star formation law (see discussion in the text).}
\label{fig-16}
\end{figure*}

Figure \ref{fig-16} shows our combined pixel-by-pixel distribution together with all the literature measurements discussed in \S\,\ref{discussion-previous}, measurements from starburst galaxies \citep{KENNICUTT98}, and data from low surface brightness galaxies (LSB, Wyder et al., in prep.). Our data and those from the literature sweep out a clear, consistent trend in SFR-gas space. The `kink' in the pixel-by-pixel distribution (indicated by the left vertical line), reflecting the saturation of \hi\ at $\Sigma_{\rm gas} \approx 9$~M$_{\odot}$~pc$^{-2}$ as discussed in \S\,\ref{saturation} and \S\,\ref{saturation-comb}, is also visible in the data distribution from the other studies.

In the following we will discuss how the results from this study relate to the work of K98, which is the only study so far to explore the SF law over the full range of $\Sigma_{\rm SFR}$ and $\Sigma_{\rm gas}$ shown in Figure \ref{fig-16}, and we will discuss a possible transition that may be expected at around $\Sigma_{\rm gas} \approx 200$ M$_{\odot}$~pc$^{-2}$ (right vertical line in Figure \ref{fig-16}).

In this paper, we find $\Sigma_{\rm SFR} \propto \Sigma_{\rm H2}^{1.0 \pm 0.2}$,
whereas K98 found a power law with slope $N = 1.40 \pm 0.15$ relating $\Sigma_{\rm SFR}$ and
$\Sigma_{\rm gas}$. The fit of K98 depends on the contrast between normal spirals, $\Sigma_{\rm H2} \approx 20$ M$_{\odot}$~pc$^{-2}$, and high surface density starbursts, $\Sigma_{\rm H2} \approx 1000$ M$_{\odot}$~pc$^{-2}$. A power law index $N \approx 1.5$ relating SFR to CO emission has been well-established in starbursts at low and high redshifts by a number of authors \citep[e.g.,][]{gao04,riechers07}. There may be reasons to expect different values of $N$ in starburst environments and in our data. Starburst galaxies have average surface densities far in excess of a Galactic GMC
\citep[e.g.,][]{gao04,rosolowsky05}. We have no such regions
in our own sample, instead we make our measurements in the regime
where $\Sigma_{\rm H2} = 3-50~$M$_{\odot}$~pc$^{-2}$. In starbursts, the changes in molecular surface density must reflect real changes in the physical conditions being observed.

In our data, $\Sigma_{\rm H2}$ is likely to be a measure of the filling factor of GMCs rather than real
variations in surface density. On the one hand, for our resolution (750 pc) and sensitivity ($\Sigma_{\rm H2} = 3$~M$_{\odot}$~pc$^{-2}$) the minimum mass we can detect along a line of sight is $\sim1.5\cdot 10^{6}$ M$_{\odot}$. Most of the mass in Galactic GMCs is in clouds with M$_{H2}\approx 5\cdot10^5-10^6$ M$_{\odot}$ \citep[e.g.,][]{blitz93}. Consequently, wherever we detect H$_{2}$ we expect at least a few GMCs in our beam. On the other hand, most of our data have $\Sigma_{\rm H2} \lesssim 50$~M$_{\odot}$~pc$^{-2}$. The typical surface density of a Galactic GMC is $170$\,M$_{\odot}$~pc$^{-2}$ \citep{solomon87}. These surface densities are much lower than those observed in starbursts and are consistent with Galactic GMCs filling $\lesssim 1/3$ of the beam. If GMC properties are the same in all spirals in our sample, then for this range of surface densities we expect a power law index of $N=1$ as $\Sigma_{\rm H2}$ just represents the beam filling fraction of GMCs. Averaging over at least a few clouds may wash out cloud-cloud variations in the SFE. A test of this interpretation is to measure GMC properties in a wide sample of spirals. We note that Local Group spirals display similar scaling relations and cloud mass distribution functions so that it is hard to distinguish GMCs in M~31 or M~33 from those in the Milky Way \citep[e.g.,][]{blitz07,bolatto08}. If this holds for all spirals, then we may indeed expect $N=1$ whenever GMCs represent the dominant mode of star formation. The next generation of mm-arrays should soon be able to measure GMC properties beyond the Local Group and shed light on this topic. In that sense, our measurement of $N=1.0 \pm 0.2$ represents a prediction that GMC properties are more or less universal in nearby spiral galaxies.

For our results to be consistent with those from starbursts, the slope must steepen near $\Sigma_{\rm H2} \approx 200$~M$_\odot$~pc$^{-2}$. This might be expected on both observational and physical grounds. CO is optically thick at the surfaces of molecular clouds. Therefore, as the filling fraction of such clouds for a given telescope beam approaches unity, CO will become an increasingly poor measure of the true $\Sigma_{\rm H2}$ because of the optical thickness of the CO emission. Even if such clouds have Galactic SFEs, the observed relationship between CO and an optically thin SFR tracer, e.g., far infrared (FIR) emission, will steepen. It is also likely that physical conditions inside the molecular gas change as surface densities exceed that of a typical GMC \citep[e.g.,][]{padoan07}. If the density increases, the free fall time within the gas will decrease, possibly leading to more efficient star formation \citep[e.g.,][]{krumholz05}. We refer the reader to \citet{krumholz07} for a full treatment of theoretical expectations for changing Schmidt laws.

In fact, \citet{gao04} may observe the transition directly; in their Figures 1b and 3 they show that FIR emission (i.e., the SFR) scales linearly with CO emission below M$_{\rm H2}\approx 10^{10}$ M$_{\odot}$, i.e., for their sample of `normal' spiral galaxies. Above $10^{10}$ M$_{\odot}$, i.e., for LIRGs and ULIRGs, they observe a steeper power law index relating SFR to CO (while the relationship between SFR and dense gas, as traced by HCN emission, remains linear). This value, M$_{\rm H2}\approx 10^{10}$ M$_{\odot}$, is about the upper limit of molecular gas masses observed in our sample of spiral galaxies \citep[see Table 5 in][]{leroy08}. Based on our results, we suggest that the former (linear) regime corresponds to star formation organized into normal spiral GMCs.

If this sketch is correct, then care must be taken relating the SFR to CO measurements.
The efficiency with which gas forms stars will depend on what regime one
considers. Perhaps more excitingly, the reverse is also true: by measuring
the SFE one can place a region or galaxy in either the `starburst' or
`GMC/disk' regime. This offers the intriguing prospect, for example, of
diagnosing the dominant mode of star formation in surveys of star
formation and molecular gas at high redshift that will be a major
component of ALMA science. Even with the current generation of
millimeter-wave telescopes such comparisons are possible. For example,
\citet{daddi08} recently showed that molecular gas in `BzK'-selected
galaxies at $z=1.5$ has the same FIR-to-CO ratio (and thus presumably SFE) as local
spiral galaxies. For comparison, sub-millimeter (`SCUBA') galaxies display
dramatically higher SFEs, as traced by their FIR-to-CO ratios \citep[e.g.,][]{greve05}. A
suggestive link to this work is that the BzK galaxies show signs of larger
stellar disks and more extended CO emission, completely consistent with
star formation proceeding mostly in a disk populated by GMCs analogous to
those in nearby galaxies.

\acknowledgments
The authors thank the teams of SINGS, the GALEX NGS and BIMA SONG for creating and making available their outstanding datasets and the anonymous referee for providing useful comments. Furthermore we would like to thank R. Kennicutt, K.~F. Schuster, T. Wong, L.~P. Crosthwaite, S. Jogee and their collaborators for making their data available. We are grateful to Rob Kennicutt and Daniela Calzetti for many useful discussions and suggestions. F.B. acknowledges support from the Deutsche Forschungsgemeinschaft (DFG) Priority Program 1177. E.B. gratefully acknowledges financial support through
an EU Marie Curie International Reintegration Grant
(Contract No. MIRG-CT-6-2005-013556). The work of W.J.G.d.B. is based upon research supported by the South African Research Chairs Initiative of the Department of Science and Technology and National Research Foundation. We have made use of the Extragalactic Database (NED), which is operated by the Jet Propulsion Laboratory, California Institute of Technology, under contract with the National Aeronautics and Space Administration. This research has made use of NASA's Astrophysics Data System (ADS).



\begin{thebibliography}{}

\bibitem[Blitz(1993)]{blitz93} Blitz, L.\ 1993, Protostars and Planets III, 125

\bibitem[Blitz et al.(2007)]{blitz07} Blitz, L., Fukui, Y., Kawamura, A., Leroy, A., Mizuno, N. \& Rosolowsky, E.\ 2007, Protostars and Planets V, 81

\bibitem[Boissier \& Prantzos(1999)]{boissier99} Boissier, S. \& Prantzos, N.\ 1999, \mnras, 307, 857

\bibitem[Boissier et al.(2003)]{boissier03} Boissier, S., Prantzos, N., Boselli, A. \& Gavazzi, G.\ 2003, \mnras, 346, 1215

\bibitem[Bolatto et al.(2008)]{bolatto08} Bolatto, A.~D., et al. 2008, \apj, accepted, astro-ph/0807.0009

\bibitem[Boselli et al.(1995)]{boselli95} Boselli, A., Gavazzi, G., Lequeux, J., Buat, V., Casoli, F., Dickey, J. \& Donas, J.\ 1995, \aap, 300, L13

\bibitem[Braine et al.(1993)]{braine93} Braine, J., Combes, F.,
Casoli, F., Dupraz, C., Gerin, M., Klein, U., Wielebinski, R. \&
Brouillet, N.\ 1993, \aaps, 97, 887

\bibitem[Buat et al.(1989)]{buat89} Buat, V., Deharveng, J.~M. \& Donas, J.\ 1989, \aap, 223, 42

\bibitem[Buat(1992)]{buat92} Buat, V.\ 1992, \aap, 264, 444

\bibitem[Calzetti et al.(2005)]{CALZETTI05} Calzetti, D., et al. 2005, \apj, 633, 871

\bibitem[Calzetti et al.(2007)]{CALZETTI07} Calzetti, D., et al.\
2007, \apj, 666, 870

\bibitem[Crosthwaite \& Turner(2007)]{crosthwaite07} Crosthwaite, L.~P., \& Turner,
J.~L.\ 2007, \aj, 134, 1827

\bibitem[Daddi et al.(2008)]{daddi08} Daddi, E., Dannerbauer, H.,
Elbaz, D., Dickinson, M., Morrison, G., Stern, D.
\& Ravindranath, S.\ 2008, \apjl, 673, L21

\bibitem[Dame et al.(2001)]{dame01} Dame, T.~M., Hartmann, D. \& Thaddeus, P.\ 2001, \apj, 547, 792

\bibitem[Deharveng et al.(1994)]{deharveng94} Deharveng, J.-M., Sasseen, T.~P., Buat, V., Bowyer, S., Lampton, M. \& Wu, X.\ 1994, \aap, 289, 715

\bibitem[de Vaucouleurs et al.(1991)]{deVaucouleurs91} de~Vaucouleurs, G., et al. 1991, Springer, Volume 1-3, XII, 2069

\bibitem[Federman et al.(1979)]{federman79} Federman, S.~R., Glassgold, A.~E. \& Kwan, J.\ 1979, \apj, 227, 466

\bibitem[Gao \& Solomon(2004)]{gao04} Gao, Y. \& Solomon, P.~M.\ 2004, \apj, 606, 271

\bibitem[Gil de Paz et al.(2007)]{gildepaz07} Gil de Paz, A., et al.\ 2007, \apjs, 173, 185

\bibitem[Gordon et al.(2005)]{gordon05} Gordon, K.~D., et al. 2005, \pasp, 177, 503

\bibitem[Greve et al.(2005)]{greve05} Greve, T.~R., et al.\ 2005, \mnras, 359, 1165

\bibitem[Hamajima \& Tosa(1975)]{hamajima75} Hamajima, K. \&
Tosa, M.\ 1975, \pasj, 27, 561

\bibitem[Hartwick(1971)]{hartwick71} Hartwick, F.~D.~A.\ 1971, \apj, 163, 431

\bibitem[Helfer et al.(2003)]{helfer03} Helfer, T.~T., Thornley, M.~D., Regan,
  M.~W., Wong, T., Sheth, K., Vogel, S.~N., Blitz, L. \& Bock, D.~C.-J.\
  2003, \apjs, 145, 259

\bibitem[Heyer et al.(2004)]{heyer04} Heyer, M.~H., Corbelli, E., Schneider, S.~E. \& Young, J.~S.\ 2004, \apj, 602, 723

\bibitem[Hunter et al.(1998)]{hunter98} Hunter, D.~A., Elmegreen, B.~G. \&
  Baker, A.~L.\ 1998, \apj, 493, 595

\bibitem[Isobe et al.(1990)]{isobe90} Isobe, T., Feigelson,
E.~D., Akritas, M.~G. \& Babu, G.~J.\ 1990, \apj, 364, 104

\bibitem[Israel(1997)]{israel97} Israel, F.~P.\ 1997, \aap, 328, 471

\bibitem[Jogee et al.(2005)]{jogee05} Jogee, S., Scoville, N.,
\& Kenney, J.~D.~P.\ 2005, \apj, 630, 837

\bibitem[J\"ors\"ater \& van\,Moorsel(1995)]{JOERSAETER95} J\"ors\"ater, S. \& van\,Moorsel, G.\,A. 1995, \aj, 110, 2037

\bibitem[Karachentsev et al.(2002)]{karachentsev02} Karachentsev, I.~D., et al. 2002, \aap, 383, 125
\bibitem[Karachentsev et al.(2003)]{karachentsev03} Karachentsev, I.~D., et al. 2003, \aap, 398, 479

\bibitem[Kennicutt(1989)]{KENNICUTT89} Kennicutt, R.~C. 1989, \apj, 344, 685

\bibitem[Kennicutt(1998a)]{KENNICUTT98} Kennicutt, R.~C. 1998a, \apj,
498, 541

\bibitem[Kennicutt(1998b)]{kennicutt-rev98} Kennicutt, R.~C., 1998b, ARA\&A, 36, 189

\bibitem[Kennicutt et al.(2003)]{KENNICUTT03} Kennicutt, R.~C., Jr., et al. 2003, \pasp, 115, 928

\bibitem[Kennicutt et al.(2007)]{KENNICUTT07} Kennicutt, R.~C.,
Jr., et al.\ 2007, \apj, 671, 333

\bibitem[Kroupa(2001)]{kroupa01} Kroupa, P., 2001, \mnras, 322, 231

\bibitem[Krumholz \& McKee(2005)]{krumholz05} Krumholz, M.~R. \& McKee,
  C.~F.\ 2005, \apj, 630, 250

\bibitem[Krumholz \& Thompson(2007)]{krumholz07} Krumholz, M.~R. \& Thompson,
  T.~A.\ 2007, \apj, 669, 289

\bibitem[Kuno et al.(2007)]{kuno07} Kuno, N., et al.\ 2007, \pasj, 59, 117

\bibitem[Lee(2006)]{lee06} Lee, J.~C., 2006, PhD Thesis, University of Arizona

\bibitem[Leitherer et al.(1999)]{leitherer99} Leitherer, C., et al.\ 1999, \apjs, 123, 3

\bibitem[Leroy et al.(2005)]{leroy05} Leroy, A., Bolatto, A.~D., Simon, J.~D.
  \& Blitz, L.\ 2005, \apj, 625, 763

\bibitem[Leroy et al.(2007)]{leroy07} Leroy, A., Bolatto, A.,
Stanimirovic, S., Mizuno, N., Israel, F., \& Bot, C.\ 2007, \apj, 658, 1027

\bibitem[Leroy et al.(2008a)]{leroy08} Leroy, A., et al. 2008a, \aj, accepted
\bibitem[Leroy et al.(2008b)]{leroy08-2} Leroy, A., et al. 2008b, \aj, submitted

\bibitem[Madden et al.(1997)]{madden97} Madden, S.~C.,
Poglitsch, A., Geis, N., Stacey, G.~J., \& Townes, C.~H.\ 1997, \apj, 483, 200

\bibitem[Madore et al.(1974)]{madore74} Madore, B.~F., van den
Bergh, S. \& Rogstad, D.~H.\ 1974, \apj, 191, 317

\bibitem[Madore(1977)]{madore77} Madore, B.~F.\ 1977, \mnras, 178, 1

\bibitem[Martin \& Kennicutt(2001)]{martin01} Martin, C.~L. \& Kennicutt, R.~C. 2001, \apj, 555, 301

\bibitem[Matteucci et al.(2006)]{matteucci06} Matteucci, F., Panagia, N., Pipino, A., Mannucci, F., Recchi, S. \& Della Valle, M.\ 2006, \mnras, 372, 265

\bibitem[Misiriotis et al.(2006)]{misiriotis06} Misiriotis, A., Xilouris, E.~M., Papamastorakis, J., Boumis, P. \& Goudis, C.~D.\ 2006, \aap, 459, 113

\bibitem[Morris \& Lo(1978)]{morris78} Morris, M., \& Lo, K.~Y.\ 1978, \apj, 223, 803

\bibitem[Morrissey et al.(2005)]{morrissey05} Morrissey, P., et al. 2005, \apj, 619, L7

\bibitem[Moustakas et al.(2006)]{moustakas06} Moustakas, J., et al. 2006, \apj, 651, 155

\bibitem[Newton(1980)]{newton80} Newton, K.\ 1980, \mnras, 190, 689

\bibitem[Padoan et al.(2007)]{padoan07} Padoan, P., Nordlund,
{\AA}., Kritsuk, A.~G., Norman, M.~L., \& Li, P.~S.\ 2007, \apj, 661, 972

\bibitem[P\'erez-Gonz\'alez et al.(2006)]{PEREZGONZALEZ06}
  P\'erez-Gonz\'alez, P.~G., et al. 2006, \apj, 648, 987

\bibitem[Prugniel \& Heraudeau(1998)]{prugniel98} Prugniel, P. \& Heraudeau, P. 1998, \aaps, 128, 299

\bibitem[Riechers et al.(2007)]{riechers07} Riechers, D.~A., Walter, F., Carilli, C.~L. \& Bertoldi, F.\ 2007, \apjl, 671, L13

\bibitem[Rieke et al.(2004)]{rieke04} Rieke, G., et al. 2004, \apjs, 154, 25

\bibitem[Rosolowsky \& Blitz(2005)]{rosolowsky05} Rosolowsky, E. \& Blitz, L.\ 2005, \apj, 623, 826

\bibitem[Salim et al.(2007)]{salim07} Salim, S., et al. 2007, \apjs, 173, 267

\bibitem[Salpeter(1955)]{salpeter55} Salpeter, E.~E., 1955, \apj, 121, 161

\bibitem[Sanduleak(1969)]{sanduleak69} Sanduleak, N.\ 1969, \aj, 74, 47

\bibitem[Schaye(2004)]{schaye04} Schaye, J. 2004, \apj, 609, 667

\bibitem[Schlegel, Finkbeiner \& Davis(1998)]{schlegel98} Schlegel, D.~J., Finkbeiner, D.~P. \& Davis, M. 1998, \apj, 500, 525

\bibitem[Schmidt(1959)]{SCHMIDT59} Schmidt, M. 1959, \apj, 129, 243

\bibitem[Schuster et al.(2004)]{schuster2004} Schuster, K.~F., et al.\ 2004,
\aap, 423, 1171

\bibitem[Schuster et al.(2007)]{schuster07} Schuster, K.~F., Kramer, C., Hitschfeld, M., Garcia-Burillo, S. \& Mookerjea, B.\ 2007, \aap, 461, 143

\bibitem[Skaya \& Federman(1987)]{skaya87} Skaya, E.~J. \& Federman, S.~R. \ 1987, \apj, 319, 76

\bibitem[Solomon et al.(1987)]{solomon87} Solomon, P.~M., Rivolo, A.~R., Barrett, J. \& Yahil, A.\ 1987, \apj, 319, 730

\bibitem[Spoon(2003)]{spoon03} Spoon, H.~W.~W.\ 2003, Ph.D.~Thesis

\bibitem[Springel \& Hernquist(2003)]{springel03} Springel, V. \& Hernquist, L.\ 2003, \mnras, 339, 289

\bibitem[Strong \& Mattox(1996)]{strong96} Strong, A.~W. \& Mattox, J.~R.\ 1996, \aap, 308, L21

\bibitem[Tamburro et al.(2008)]{tamburro08} Tamburro, D., et al. 2008, \aj, accepted

\bibitem[Tan et al.(1999)]{tan99} Tan, J.~C., Silk, J. \& Balland, C.\ 1999, \apj, 522, 579

\bibitem[Taylor et al.(1998)]{taylor98} Taylor, C.~L.,
Kobulnicky, H.~A. \& Skillman, E.~D.\ 1998, \aj, 116, 2746

\bibitem[Thilker et al.(2005)]{thilker05} Thilker, D.~A., et al.\ 2005, \apjl, 619, L79

\bibitem[Thornley \& Wilson(1995)]{thornley95} Thornley, M.~D. \&
Wilson, C.~D.\ 1995, \apj, 447, 616

\bibitem[Tosa \& Hamajima(1975)]{tosa75} Tosa, M. \&
Hamajima, K.\ 1975, \pasj, 27, 501

\bibitem[Walter \& Brinks(1999)]{WALTER99} Walter, F. \& Brinks, E. 1999, \aj, 118, 273

\bibitem[Walter et al.(2001)]{walter01} Walter, F., Taylor, C.~L.,
  H{\"u}ttemeister, S., Scoville, N. \& McIntyre, V.\ 2001, \aj, 121, 727

\bibitem[Walter et al.(2008)]{WALTER08} Walter, F. 2008, \aj, accepted

\bibitem[Wilson(1995)]{wilson95} Wilson, C.~D.\ 1995, \apjl, 448, L97

\bibitem[Wong \& Blitz(2002)]{wong02} Wong, T. \& Blitz, L. 2002, \apj, 569, 157

\bibitem[Wyder et al.(2007)]{wyder07} Wyder, T., et al. 2007, \apjs, 173, 293

\bibitem[Young et al.(1995)]{young95} Young, J.~S., et al. 1995, \apjs, 98, 219

\end{thebibliography}
\end{document}